\documentclass[11pt,a4paper]{article}

\usepackage[utf8]{inputenc}
\usepackage{amsmath,latexsym}
\usepackage{mathrsfs}
\usepackage{mathtools}
\usepackage[dvipsnames,table]{xcolor} 
\usepackage{graphicx}
\usepackage{slashed}

\usepackage{tikz}
\usepackage[compat=1.0.0]{tikz-feynman}
\usepackage[nomessages]{fp}
\usepackage{jheppub}
\usepackage{graphicx}
\usepackage{soul}

\usepackage{changepage} 
\usepackage{braket}
\usepackage{relsize}
\usepackage{lscape} 
\usepackage{bbm} 
\usepackage{pdfpages} 
\newcommand{\minus}{\scalebox{0.75}[1.0]{$-$}}
\usepackage{nicematrix}
\usepackage{array,xparse,l3keys2e,expl3}
\usepackage{float}
\usepackage{multicol}
\usepackage{scalerel}

\usepackage{multirow}

\def\subN{{\scaleto{N}{5pt}}}
\def\subM{{\scaleto{M}{5pt}}}
\def\subsubN{{\scaleto{N}{4pt}}}

\def\beq{\begin{equation}}   
\def\eeq{\end{equation}}
\def\bea{\begin{eqnarray}}  
\def\eea{\end{eqnarray}} 
\def\nn{\nonumber}
\def\r{\right} 
\def\l{\left} 

\newtheorem{thm}{Theorem}[section]
\newtheorem{cor}[thm]{Corollary}
\newtheorem{lem}[thm]{Lemma}

\def\eps{\varepsilon}
\def\O{\mathcal{O}}
\def\L{\mathcal{L}}
\def\Z{\mathcal{Z}}

\def\p{\phi}
\def\d{\partial}

\usepackage[normalem]{ulem}

\definecolor{darkgreen}{rgb}{0,0.5,0}

\title{Renormalization and non-renormalization theorems  with multiple operator insertions}
\title{Non-linear non-renormalization theorems}

\author[a,b]{Weiguang Cao,}
\author[c]{Franz Herzog,}
\author[a]{Tom Melia,}
\author[d]{and Jasper Roosmale Nepveu}

\affiliation[a]{Kavli Institute for the Physics and Mathematics of the Universe (WPI),\\ The University of Tokyo Institutes for Advanced Study, The University of Tokyo, Kashiwa, Chiba 277-8583, Japan}
\affiliation[b]{Department of Physics, Graduate School of Science, The University of Tokyo, Tokyo 113-0033, Japan}
\affiliation[c]{Higgs Centre for Theoretical Physics, School of Physics and Astronomy, The University of Edinburgh, Edinburgh EH9 3FD, Scotland, UK}
\affiliation[d]{Humboldt-Universit\"at zu Berlin,\\ Institut f\"ur Physik, D-12489 Berlin,
Germany}

\emailAdd{weiguang.cao@ipmu.jp, fherzog@ed.ac.uk, tom.melia@ipmu.jp, jasper.roosmalenepveu@physik.hu-berlin.de}

\abstract{
We study the mixing of operators under renormalization group flow in quantum theories, and
prove a non-renormalization theorem at non-linear order.
It dictates zeros up to a certain  number of loops in anomalous dimension tensors that control, for example, the mixing of operators at order dimension six squared into dimension eight. We obtain new results at up to three loops
 for the mass dimension eight anomalous dimension tensor of $\phi^4$ theory in $D=4-2\eps$ dimensions and verify the zeros predicted by the theorem. 
}

\begin{document}

\begin{flushright}
HU-EP-23/06-RTG
\vspace{-0.9cm}
\end{flushright}
\flushbottom

\maketitle
\newpage

\section{Introduction}

Quantum field theory is notorious for harbouring deep and often surprising structures. The Parke-Taylor formulae~\cite{Parke:1986gb} are paradigmatic of an underlying simplicity that exists in gauge theory scattering amplitudes. 
BCJ relations and the double copy formulation of gauge and gravity theories~\cite{Bern:2008qj,Bern:2010ue}, and a geometric amplituhedra description~\cite{Arkani-Hamed:2013jha} all manifest structure that is completely obscured in the Lagrangian description.
More recently the extent to which these structures are present in effective field theory (EFT) has been explored. On-shell methodologies and recursion relations~\cite{Britto:2005fq} have  been applied to EFT amplitudes~\cite{Cheung:2014dqa,Cheung:2015ota,Cheung:2016drk,Henning:2017fpj,Shadmi:2018xan,Henning:2019enq,Henning:2019mcv,Ma:2019gtx,Durieux:2019eor,Aoude:2019tzn,Durieux:2019siw}; a version of the double copy relates amplitudes of different EFTs~\cite{
Broedel:2012rc,
Elvang:2018dco, 
CarrilloGonzalez:2019fzc, 
Carrasco:2019yyn,
Carrasco:2021ptp,  
Chi:2021mio, 
Bonnefoy:2021qgu,
Carrasco:2022jxn,Carrasco:2022lbm,Chen:2023dcx} (for both scalar and gauge EFTs);
and the counterparts of amplituhedra have been explored~\cite{Arkani-Hamed:2020gyp, Arkani-Hamed:2020blm}.

Another object in which an unexpected structure in EFT has been observed is the anomalous dimension matrix (ADM) that describes the mixing of operators into each other under the renormalization group flow at linear order. Even in a theory as complicated as the Standard Model (SM) EFT one observes the appearance of zeros and approximate holomorphy~\cite{Alonso:2014rga}; theorems regarding such zeros for the ADM were proven at one-loop order~\cite{Elias-Miro:2014eia,Cheung:2015aba}, and general theorem for single operator insertions was derived in \cite{Bern:2019wie}, which yields zeros at higher loop orders for theories not containing relevant couplings. Further non-renomalization theorems have also been discussed in generic EFTs in \cite{Jiang:2020rwz}, based on angular momentum selection rules, and in gravity \cite{Baratella:2021guc}. Their effect on tree-level and one-loop amplitudes was examined in \cite{Craig:2019wmo}. Also in a pure scalar EFT, the ADM exhibits further one and three loop zeros in the operator basis of conformal primaries \cite{Cao:2021cdt}.

\begin{figure}
\centering
\includegraphics[clip, trim=1cm 10cm 10cm 1cm, width=10cm]{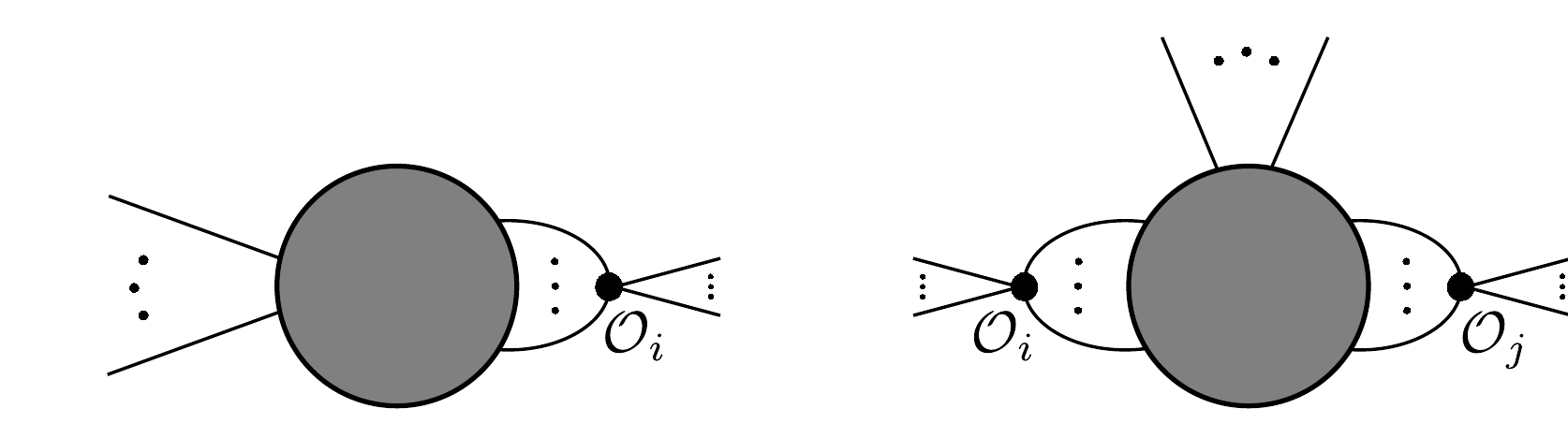}
\caption{Left: insertion of a higher-dimension operator into an amplitude at linear order. Right: insertion of two higher dimension operators into an amplitude, leading to non-linear renormalization}
\label{fig:one}
\end{figure}

At higher order in the EFT expansion, renormalization also requires non-linear mixing of operators. For example, operators of mass dimension six---inserted twice in a diagram---can renormalize dimension-eight operators; see Fig.~\ref{fig:one}. This  mixing is controlled by the anomalous dimension {\it tensor} (ADT). In  the SM, the first calculations at the one-loop level of ADTs were presented in~\cite{Jenkins:2017dyc, Davidson:2018zuo,Chala:2021juk, Chala:2021pll,Chala:2021wpj,Silva:2022tln,Helset:2022pde,Asteriadis:2022ras}.
Besides being theoretically interesting, higher order effects in the EFT expansion can also be important phenomenologically \cite{Dawson:2021xei}. For instance mass dimension eight operators may dominate over dimension six ones~\cite{Azatov:2016sqh}. Dimension six squared effects were also studied in \cite{Kim:2022amu} and the impact of quadratic terms on the error estimate was recently discussed in \cite{Martin:2021vwf,Trott:2021vqa}.

Similar to the non-renormalization theorems of ADMs, the zeros in the ADT have interesting and unexpected  patterns. 
In this work, we provide the first non-renormalization theorem beyond the linear order, valid for general EFTs without relevant couplings or masses in the Lagrangian. The theorem provides the `most basic' lower bounds on the loop order to which certain entries in the ADT must remain zero. Using on-shell amplitude and Feynman graph arguments, our proof of the theorem reproduces the same lower bound as~\cite{Bern:2019wie} when applied to the case of one insertion (i.e.\ the linear order ADM). EFTs with a given particle spectrum may display further selection rules such that zeros may persist to higher loop order than the bounds we give. We leave a study of this to future work, but note that our bounds are minimal in the sense that they still provide a lower bound for general theories.

As well as deriving a non-renormalization theorem, we also provide results for the ADT of scalar EFTs at higher loop order. Specifically, we give results for the ADT at mass dimension eight (i.e.\ dimension six squared) up to three loops for both real and complex scalar EFT. These results go beyond what is in the literature in terms of {both} mass dimension and loop-order. To obtain these results we apply calculational technologies developed in a previous publication~\cite{Cao:2021cdt}, namely the $R^*$ methodology~\cite{Chetyrkin:1982nn,Chetyrkin:1984xa,Smirnov:1986me,Chetyrkin:2017ppe}, based on a formulation valid for general QFTs~\cite{Herzog:2017bjx,deVries:2019nsu,Beekveldt:2020kzk}, combined with a use of Hilbert series and conformal primary bases in EFT, see~\cite{Henning:2015daa,Lehman:2015via,Lehman:2015coa,Henning:2015alf,Henning:2017fpj,Kobach:2017xkw,Kobach:2018nmt,Kobach:2018pie,Henning:2019enq,Henning:2019mcv,Ruhdorfer:2019qmk,Graf:2020yxt,Graf:2022rco}. We verify the zeros found in these ADTs are explained by our theorem. 
These non-linear results are also relevant for the computation of conformal field theory (CFT) data of $\p^4$ theory at the Wilson-Fisher fixed point. 

The paper proceeds as follows. In Section~\ref{s:background} we review  renormalization theory in EFT, which serves to provide the conventions we use throughout. In Section~\ref{s:structure} we prove the non-renormalization theorem for the ADT. Section~\ref{s:results} provides results for the mass dimension eight ADT up to three loops in scalar EFT. Section~\ref{s:disc} concludes.

\section{Background: renormalization beyond single insertions}
\label{s:background}
To set the conventions for the rest of this paper,
we first review the renormalization theory of the massless $Z_2$-symmetric real scalar EFT in four dimensions, with emphasis on beyond linear order in the Wilson coefficients. 
This discussion extends straightforwardly to the massless complex scalar EFT, and more general theories. 
A more detailed treatment restricted to single insertions of higher dimension operators can be found in \cite{Cao:2021cdt}.
Section \ref{s:eftlagr} defines the off-shell and physical bases used in the calculation of our results in Sec.\ \ref{s:results}. Sections \ref{s:ADToffshell} and \ref{s:ADTFF} discuss the appearance of counterterms in off-shell and on-shell calculations, respectively.

\subsection{EFT Lagrangian and operator bases}\label{s:eftlagr}

The Lagrangian of the massless $Z_2$-symmetric real scalar EFT is 
\beq
\L=\L^{(4)}(\phi,\d_\mu\p)+\sum_{n>4}
 \frac{1}{\Lambda^{n-4}}
\L^{(n)}(\phi,\d_\mu\p,\d_{\mu}\d_\nu\p,...)\,,
\eeq
where 
\beq
\L^{(4)}(\phi,\d_\mu\p)=\frac{1}{2}(\d_\mu\p^b)(\d^\mu\p^b)-g^b \frac{1}{4!} (\p^b)^4\, ,
\eeq
and 
\beq
\L^{(n)}(\phi,\d_\mu\p,\d_{\mu}\d_\nu\p,...)=\sum_i \widetilde c^{\, b(n)}_i \, \widetilde \O_i^{b(n)} \, .
\label{eq:local}
\eeq
In the above, $\Lambda$ is a high energy scale which serves as the expansion parameter in the EFT, and in Eq.\ \eqref{eq:local} the Lagrangian  at mass dimension $n$ is expanded in terms of bare local operators $\widetilde \O_i^{b(n)}$  and their corresponding bare Wilson coefficients $\widetilde c^{\, b(n)}_i$. 
In our calculations, we only consider the space-time parity-even sector, but the non-renormalization theorem presented in Sec.\ \ref{s:structure} holds more generally.
The significance of the tildes in the expression will be explained in the subsequent paragraphs. Throughout, the superscript $b$ indicates a bare quantity (coupling or field) which is related to its renormalized counterpart in dimensional regularization ($D=4 - 2\,\eps$ dimensions),
\beq
\phi^b=\sqrt{Z_2}\,\phi\,,\qquad g^b={(4\pi)^2} \, {Z_g \, g(\mu) \, \mu^{2\eps}}.
\eeq
The operators $\widetilde \O_i^{b(n)}$ are polynomial in the bare field $\p^b$ and its derivatives and can therefore be renormalized as follows:
\beq \label{bareoperators}
\widetilde \O_i^{b(n)}=
\big(g^b\big)^{(\ell_i-2)/2} \, Z_2^{\, \ell_i/2}
\,\widetilde \O_i^{(n)}\,,
\eeq
where $\ell_i$ is the number of fields, or the \emph{length}, of the operator $\widetilde \O_i^{(n)}$.
In the above definition, we have included a factor of the `renormalizable' coupling $g^b$ for convenience, as will become clear later.

Let us now return to the meaning of the tildes on the operators and couplings in the above equations. In general, an EFT Lagrangian may contain redundant parameters due to relations between operators through integration by parts (IBP) and field redefinitions.
At the above Lagrangian stage, the considered set of operators $\big\{\widetilde \O_i^{(n)}\big\}$ is assumed independent under IBP only. 
These operators and their couplings are written with a tilde to indicate that they span an \emph{off-shell basis}.%
    \footnote{An off-shell basis is sometimes called a Green's basis in the literature \cite{Jiang:2018pbd}. }
That is, they constitute the counterterms required to renormalize all off-shell Green's functions.

When it comes to physical (on-shell) S-matrix elements, which are invariant under field redefinitions, the off-shell basis is overcomplete.  
An independent basis for the S-matrix will be called a \emph{physical basis}, and the corresponding operators and couplings will be written without tildes. 
A convenient physical basis is given by a subset of the off-shell basis in which no linear combination of operators vanishes on shell. Such bases will in the following be referred to as \emph{on-shell bases}.
The operators in an on-shell basis are in one-to-one correspondence to the (minimal) contact-term contributions to on-shell amplitudes (which is why such bases are also referred to as amplitude bases in the literature, see e.g.\ \cite{Ma:2019gtx}).
In the scalar EFTs, on-shell bases can be obtained from an off-shell basis by removing any operator that is proportional to $\partial^2 \phi$ (i.e.\ containing self-loops in the graphical notation explained below),  after application of integration by parts if necessary.

Within the context of $\phi^4$ theory, an off-shell basis can be obtained through a correspondence between multigraphs and operators: fields are represented by dots in the multigraph and the contractions of derivatives are represented by lines between those dots (more than one connecting line is allowed, hence the term multigraph). One choice of off-shell basis is given by operators that correspond to all multigraphs without self loops \cite{Cao:2021cdt}. 
For example, at mass dimension 6,
\begin{align}
\label{eq:scalar operators}
	\mathcal{\widetilde{O}}^{\scaleto{(6)}{7pt}}_{6} &=  \frac{1}{6!} \p^6= \frac{1}{6!} \hspace{1mm}
			\begin{gathered}
			\begin{tikzpicture}	
			\begin{feynman}[small, baseline=g1]
					\tikzfeynmanset{every vertex={dot,black,minimum size=1mm}}
				\vertex  (g1);
				\vertex [right =0.3cm of g1] (g2) ;
				\vertex [below =0.3cm of g1] (g3);
				\vertex [right =0.3cm of g3] (g4);
				\vertex [below =0.3cm of g3] (g5);
				\vertex [below =0.3cm of g4] (g6);
				\diagram* {
				};	
			\end{feynman}
			\end{tikzpicture}
			\end{gathered}\hspace{1mm}
			\,,\\[1mm]
			\mathcal{\widetilde{O}}^{\scaleto{(6)}{7pt}}_4 &=\minus\frac{1}{4} \p^2\d^{\mu}\p\d_{\mu}\p=\minus\frac{1}{4} \hspace{1mm}
			\begin{gathered}
			\begin{tikzpicture}	
			\begin{feynman}[small, baseline=g1]
					\tikzfeynmanset{every vertex={dot,black,minimum size=1mm}}
				\vertex  (g1);
				\vertex [right =0.3cm of g1] (g2) ;
				\vertex [below =0.3cm of g1] (g3);
				\vertex [right =0.3cm of g3] (g4);
				\diagram* {
					(g1) -- [] (g2)
				};	
			\end{feynman}
			\end{tikzpicture}
			\end{gathered}\hspace{1mm} 
			\,,\\[1mm]
			\mathcal{\widetilde{O}}^{\scaleto{(6)}{7pt}}_2 &= \frac{1}{2}\d^{\mu}\d^{\nu}\p\d_{\mu}\d_{\nu}\p= \frac{1}{2} \hspace{1mm} 
			\begin{gathered}
			\begin{tikzpicture}	
			\begin{feynman}[small, baseline=g1]
					\tikzfeynmanset{every vertex={dot,black,minimum size=1mm}}
				\vertex  (g1);
				\vertex [right =0.3cm of g1] (g2) ;
				\diagram* {
					(g1) -- [out=30,in=150] (g2) -- [out=-150,in=-30] (g1)
				};	
			\end{feynman}
			\end{tikzpicture}
			\end{gathered}
			\,.
\end{align}
In Table \ref{table:1} we introduce a number of operators in the multigraph representation which are used in explicit calculations later on. For the complex scalar, we use black and white dots to differentiate between $\phi$ and $\phi^\dagger$ fields.

Regarding the physical basis, it is in general a non-trivial exercise to remove both the IBP and field-redefinition redundancies directly at the Lagrangian level. However, in recent years, Hilbert series techniques have successfully been employed to enumerate the number of operators in a physical basis by utilizing conformal representation theory and commutative algebra \cite{Henning:2015daa,Henning:2015alf,Henning:2017fpj,Henning:2019mcv,Graf:2020yxt,Cao:2021cdt,Graf:2022rco}. In particular it was elucidated that 
the set of primary, scalar operators form a choice of physical basis (and this is an example of an on-shell basis).
The primary operators are thus identified as a special choice of physical basis, and additional zeros were found in the ADM with this choice \cite{Cao:2021cdt}. 
We present our results in section \ref{s:results} in the basis of conformal primaries. This basis is defined at mass dimension six and eight for both the real and the complex scalar theories in table \ref{table:2}. 

\begin{table}\centering
\begin{tabular}{ |l|l| } 
\hline
	 \textbf{Real scalar theory }& \textbf{Complex scalar theory} \\
\hline
&\\[-2mm]
$ \mathcal{O}^{\scaleto{(6)}{7pt}}_{\scaleto{6}{5pt}} \hspace{2mm} = \hspace{2mm} \frac{1}{6!} \hspace{2mm} 
			\begin{gathered}
			\begin{tikzpicture}	
			\begin{feynman}[small, baseline=g1]
					\tikzfeynmanset{every vertex={dot,black,minimum size=1mm}}
				\vertex  (g1);
				\vertex [right =0.3cm of g1] (g2) ;
				\vertex [below =0.3cm of g1] (g3);
				\vertex [right =0.3cm of g3] (g4);
				\vertex [below =0.3cm of g3] (g5);
				\vertex [below =0.3cm of g4] (g6);
				\diagram* {
				};	
			\end{feynman}
			\end{tikzpicture}
			\end{gathered} $
& $\mathcal{O}^{\scaleto{(6)}{7pt}}_{\scaleto{6}{5pt}} \hspace{2mm} = \hspace{2mm} \frac{1}{3!3!} \hspace{2mm} 
			\begin{gathered}
			\begin{tikzpicture}	
			\begin{feynman}[small, baseline=g1]
					\tikzfeynmanset{every vertex={dot,black,minimum size=1mm}}
				\vertex  (g1);
				\vertex [below =0.3cm of g1] (g2) ;
				\vertex [below =0.3cm of g2] (g3);
					\tikzfeynmanset{every vertex={empty dot,black,minimum size=1mm}}
				\vertex [right =0.3cm of g1] (g4);
				\vertex [below =0.3cm of g4] (g5);
				\vertex [below =0.3cm of g5] (g6);
				\diagram* {
				};	
			\end{feynman}
			\end{tikzpicture}
			\end{gathered}$ 
			\\[3mm]
 $\mathcal{O}^{\scaleto{(6)}{7pt}}_4 \hspace{2mm} = \hspace{2mm} \minus\frac{1}{4} \hspace{2mm} 
			\begin{gathered}
			\begin{tikzpicture}	
			\begin{feynman}[small, baseline=g1]
					\tikzfeynmanset{every vertex={dot,black,minimum size=1mm}}
				\vertex  (g1);
				\vertex [right =0.3cm of g1] (g2) ;
				\vertex [below =0.3cm of g1] (g3);
				\vertex [right =0.3cm of g3] (g4);
				\diagram* {
					(g1) -- [] (g2)
				};	
			\end{feynman}
			\end{tikzpicture}
			\end{gathered}$ 
& $\mathcal{O}^{\scaleto{(6)}{7pt}}_{4,i} \hspace{2mm} = \hspace{2mm} \minus \hspace{2mm} 
			\begin{gathered}
			\begin{tikzpicture}	
			\begin{feynman}[small, baseline=g1]
					\tikzfeynmanset{every vertex={dot,black,minimum size=1mm}}
				\vertex  (g1);
				\vertex [below =0.3cm of g1] (g3);
					\tikzfeynmanset{every vertex={empty dot,black,minimum size=1mm}}
				\vertex [right =0.3cm of g1] (g2) ;
				\vertex [right =0.3cm of g3] (g4);
				\diagram* {
					(g1) -- [] (g2)
				};	
			\end{feynman}
			\end{tikzpicture}
			\end{gathered}
\hspace{2mm} , \hspace{2mm} \minus\frac{1}{4} \hspace{2mm}
			\begin{gathered}
			\begin{tikzpicture}	
			\begin{feynman}[small, baseline=g1]
					\tikzfeynmanset{every vertex={dot,black,minimum size=1mm}}
				\vertex  (g1);
				\vertex [below =0.3cm of g1] (g3);
					\tikzfeynmanset{every vertex={empty dot,black,minimum size=1mm}}
				\vertex [right =0.3cm of g1] (g2) ;
				\vertex [right =0.3cm of g3] (g4);
				\diagram* {
					(g1) -- [] (g3)
				};	
			\end{feynman}
			\end{tikzpicture}
			\end{gathered}
\hspace{2mm} , \hspace{2mm} \minus\frac{1}{4} \hspace{2mm}
			\begin{gathered}
			\begin{tikzpicture}	
			\begin{feynman}[small, baseline=g1]
					\tikzfeynmanset{every vertex={dot,black,minimum size=1mm}}
				\vertex  (g1);
				\vertex [below =0.3cm of g1] (g3);
					\tikzfeynmanset{every vertex={empty dot,black,minimum size=1mm}}
				\vertex [right =0.3cm of g1] (g2) ;
				\vertex [right =0.3cm of g3] (g4);
				\diagram* {
					(g2) -- [] (g4)
				};	
			\end{feynman}
			\end{tikzpicture}
			\end{gathered}$ 
			\\[2mm] 
 $\mathcal{O}^{\scaleto{(6)}{7pt}}_2 \hspace{2mm} = \hspace{2mm} \frac{1}{2} \hspace{2mm} 
			\begin{gathered}
			\begin{tikzpicture}	
			\begin{feynman}[small, baseline=g1]
					\tikzfeynmanset{every vertex={dot,black,minimum size=1mm}}
				\vertex  (g1);
				\vertex [right =0.3cm of g1] (g2) ;
				\diagram* {
					(g1) -- [out=30,in=150] (g2) -- [out=-150,in=-30] (g1)
				};	
			\end{feynman}
			\end{tikzpicture}
			\end{gathered}$ 
& $\mathcal{O}^{\scaleto{(6)}{7pt}}_{2} \hspace{2mm} = \hspace{2mm} \hspace{2mm} 
			\begin{gathered}
			\begin{tikzpicture}	
			\begin{feynman}[small, baseline=g1]
					\tikzfeynmanset{every vertex={dot,black,minimum size=1mm}}
				\vertex  (g1);
					\tikzfeynmanset{every vertex={empty dot,black,minimum size=1mm}}
				\vertex [right =0.3cm of g1] (g2) ;
				\diagram* {
					(g1) -- [out=30,in=150] (g2) -- [out=-150,in=-30] (g1)
				};	
			\end{feynman}
			\end{tikzpicture}
			\end{gathered}$ \\[2mm]
\hline
&\\[-2mm]
$\mathcal{O}^{\scaleto{(8)}{7pt}}_{\scaleto{8}{5pt}} \hspace{2mm} = \hspace{2mm} \frac{1}{8!} \hspace{2mm} 
			\begin{gathered}
			\begin{tikzpicture}	
			\begin{feynman}[small, baseline=g1]
					\tikzfeynmanset{every vertex={dot,black,minimum size=1mm}}
				\vertex  (g1);
				\vertex [right =0.3cm of g1] (g2) ;
				\vertex [below =0.3cm of g1] (g3) ;
				\vertex [right =0.3cm of g3] (g4) ;
				\vertex [below =0.3cm of g3] (g5) ;
				\vertex [right =0.3cm of g5] (g6) ;
				\vertex [below =0.3cm of g5] (g7) ;
				\vertex [right =0.3cm of g7] (g8) ;
				\diagram* {
				};	
			\end{feynman}
			\end{tikzpicture}
			\end{gathered}$
&$\mathcal{O}^{\scaleto{(8)}{7pt}}_{\scaleto{8}{5pt}} \hspace{2mm} = \hspace{2mm} \frac{1}{4!4!} \hspace{2mm} 
			\begin{gathered}
			\begin{tikzpicture}	
			\begin{feynman}[small, baseline=g1]
					\tikzfeynmanset{every vertex={dot,black,minimum size=1mm}}
				\vertex  (g1);
				\vertex [below =0.3cm of g1] (g2) ;
				\vertex [below =0.3cm of g2] (g3);
				\vertex [below =0.3cm of g3] (g4);
					\tikzfeynmanset{every vertex={empty dot,black,minimum size=1mm}}
				\vertex [right =0.3cm of g1] (g6);
				\vertex [below =0.3cm of g6] (g7);
				\vertex [below =0.3cm of g7] (g8);
				\vertex [below =0.3cm of g8] (g9);
				\diagram* {
				};	
			\end{feynman}
			\end{tikzpicture}
			\end{gathered}$
			\\[4mm]
$\mathcal{O}^{\scaleto{(8)}{7pt}}_{\scaleto{6}{5pt}} \hspace{2mm} = \hspace{2mm} \minus\frac{1}{48} \hspace{2mm} 
			\begin{gathered}
			\begin{tikzpicture}	
			\begin{feynman}[small, baseline=g1]
					\tikzfeynmanset{every vertex={dot,black,minimum size=1mm}}
				\vertex  (g1);
				\vertex [right =0.3cm of g1] (g2) ;
				\vertex [below =0.3cm of g1] (g3) ;
				\vertex [right =0.3cm of g3] (g4) ;
				\vertex [below =0.3cm of g3] (g5) ;
				\vertex [right =0.3cm of g5] (g6) ;
				\diagram* {
					(g1) -- [] (g2);
				};	
			\end{feynman}
			\end{tikzpicture}
			\end{gathered}$
&$\mathcal{O}^{\scaleto{(8)}{7pt}}_{6,i} \hspace{2mm} = \hspace{2mm} \minus\frac{1}{4} \hspace{2mm} 
			\begin{gathered}
			\begin{tikzpicture}	
			\begin{feynman}[small, baseline=g1]
					\tikzfeynmanset{every vertex={dot,black,minimum size=1mm}}
				\vertex  (g1);
				\vertex [below =0.3cm of g1] (g3);
				\vertex [below =0.3cm of g3] (l1);
					\tikzfeynmanset{every vertex={empty dot,black,minimum size=1mm}}
				\vertex [right =0.3cm of g1] (g2) ;
				\vertex [below =0.3cm of g2] (r1);
				\vertex [below =0.3cm of r1] (r2);
				\diagram* {
					(g1) -- [] (g2)
				};	
			\end{feynman}
			\end{tikzpicture}
			\end{gathered}
\hspace{2mm} , \hspace{2mm} \minus\frac{1}{12} \hspace{2mm}
			\begin{gathered}
			\begin{tikzpicture}	
			\begin{feynman}[small, baseline=g1]
					\tikzfeynmanset{every vertex={dot,black,minimum size=1mm}}
				\vertex  (g1);
				\vertex [below =0.3cm of g1] (g3);
				\vertex [below =0.3cm of g3] (l1);
					\tikzfeynmanset{every vertex={empty dot,black,minimum size=1mm}}
				\vertex [right =0.3cm of g1] (g2) ;
				\vertex [below =0.3cm of g2] (r1);
				\vertex [below =0.3cm of r1] (r2);
				\diagram* {
					(g1) -- [] (g3)
				};	
			\end{feynman}
			\end{tikzpicture}
			\end{gathered}
\hspace{2mm} , \hspace{2mm} \minus\frac{1}{12} \hspace{2mm}
			\begin{gathered}
			\begin{tikzpicture}	
			\begin{feynman}[small, baseline=g1]
					\tikzfeynmanset{every vertex={dot,black,minimum size=1mm}}
				\vertex  (g1);
				\vertex [below =0.3cm of g1] (g3);
				\vertex [below =0.3cm of g3] (l1);
					\tikzfeynmanset{every vertex={empty dot,black,minimum size=1mm}}
				\vertex [right =0.3cm of g1] (g2) ;
				\vertex [below =0.3cm of g2] (g4);
				\vertex [below =0.3cm of g4] (r2);
				\diagram* {
					(g2) -- [] (g4)
				};	
			\end{feynman}
			\end{tikzpicture}
			\end{gathered}$
			\\[3mm]
$\mathcal{O}^{\scaleto{(8)}{7pt}}_{4,i} \hspace{2mm} = \hspace{2mm} \frac{1}{4} \hspace{2mm} 
			\begin{gathered}
			\begin{tikzpicture}	
			\begin{feynman}[small, baseline=g1]
					\tikzfeynmanset{every vertex={dot,black,minimum size=1mm}}
				\vertex  (g1);
				\vertex [right =0.3cm of g1] (g2) ;
				\vertex [below =0.3cm of g1] (g3) ;
				\vertex [right =0.3cm of g3] (g4) ;
				\diagram* {
					(g1) -- [out=30,in=150] (g2) -- [out=-150,in=-30] (g1)
				};	
			\end{feynman}
			\end{tikzpicture}
			\end{gathered}
\hspace{2mm}	, \hspace{2mm} \frac{1}{2} \hspace{2mm} 
			\begin{gathered}
			\begin{tikzpicture}	
			\begin{feynman}[small, baseline=g1]
					\tikzfeynmanset{every vertex={dot,black,minimum size=1mm}}
				\vertex  (g1);
				\vertex [right =0.3cm of g1] (g2) ;
				\vertex [below =0.3cm of g1] (g3) ;
				\vertex [right =0.3cm of g3] (g4) ;
				\diagram* {
					(g1) -- [] (g2);
					(g1) -- [] (g3);
				};	
			\end{feynman}
			\end{tikzpicture}
			\end{gathered}
\hspace{2mm}	, \hspace{2mm} \frac{1}{8} \hspace{2mm} 
			\begin{gathered}
			\begin{tikzpicture}	
			\begin{feynman}[small, baseline=g1]
					\tikzfeynmanset{every vertex={dot,black,minimum size=1mm}}
				\vertex  (g1);
				\vertex [right =0.3cm of g1] (g2) ;
				\vertex [below =0.3cm of g1] (g3) ;
				\vertex [right =0.3cm of g3] (g4) ;
				\diagram* {
					(g1) -- [] (g2);
					(g3) -- [] (g4);
				};	
			\end{feynman}
			\end{tikzpicture}
			\end{gathered}$
&$\mathcal{O}^{\scaleto{(8)}{7pt}}_{4,i} \hspace{2mm} = \hspace{3mm} 
		\begin{gathered}
			\begin{tikzpicture}	
			\begin{feynman}[small, baseline=g1]
					\tikzfeynmanset{every vertex={dot,black,minimum size=1mm}}
				\vertex  (l1);
				\vertex [below =0.3cm of l1] (l2);
					\tikzfeynmanset{every vertex={empty dot,black,minimum size=1mm}}
				\vertex [right =0.3cm of l1] (r1) ;
				\vertex [below =0.3cm of r1] (r2);
				\diagram* {
					(l1) -- [out=30,in=150] (r1) -- [out=-150,in=-30] (l1)
				};	
			\end{feynman}
			\end{tikzpicture}
			\end{gathered}
\hspace{2mm} , \hspace{2mm} \frac{1}{4} \hspace{2mm}
		\begin{gathered}
			\begin{tikzpicture}	
			\begin{feynman}[small, baseline=g1]
					\tikzfeynmanset{every vertex={dot,black,minimum size=1mm}}
				\vertex  (l1);
				\vertex [below =0.3cm of l1] (l2);
					\tikzfeynmanset{every vertex={empty dot,black,minimum size=1mm}}
				\vertex [right =0.3cm of l1] (r1) ;
				\vertex [below =0.3cm of r1] (r2);
				\diagram* {
					(l1) -- [out=-60,in=60] (l2) -- [out=120,in=-120] (l1)
				};	
			\end{feynman}
			\end{tikzpicture}
			\end{gathered}
\hspace{2mm} , \hspace{2mm} \frac{1}{4} \hspace{2mm}
		\begin{gathered}\begin{tikzpicture} \begin{feynman}[small, baseline=g1]
					\tikzfeynmanset{every vertex={dot,black,minimum size=1mm}}
				\vertex  (l1);
				\vertex [below =0.3cm of l1] (l2);
					\tikzfeynmanset{every vertex={empty dot,black,minimum size=1mm}}
				\vertex [right =0.3cm of l1] (r1) ;
				\vertex [below =0.3cm of r1] (r2);
				\diagram* {
					(r1) -- [out=-60,in=60] (r2) -- [out=120,in=-120] (r1)
				};	
			\end{feynman}\end{tikzpicture}\end{gathered}$\ , 
			\\[2mm]
&$\hspace{16.7mm} 
%
%
		\begin{gathered}
			\begin{tikzpicture}	
			\begin{feynman}[small, baseline=g1]
					\tikzfeynmanset{every vertex={dot,black,minimum size=1mm}}
				\vertex  (l1);
				\vertex [below =0.3cm of l1] (l2);
					\tikzfeynmanset{every vertex={empty dot,black,minimum size=1mm}}
				\vertex [right =0.3cm of l1] (r1) ;
				\vertex [below =0.3cm of r1] (r2);
				\diagram* {
					(l1) -- [] (r1),
					(l1) -- [] (l2)
				};	
			\end{feynman}
			\end{tikzpicture}
			\end{gathered}
%
%
%
\hspace{2mm} , \hspace{3mm}
		\begin{gathered}\begin{tikzpicture} \begin{feynman}[small, baseline=g1]
					\tikzfeynmanset{every vertex={dot,black,minimum size=1mm}}
				\vertex  (l1);
				\vertex [below =0.3cm of l1] (l2);
					\tikzfeynmanset{every vertex={empty dot,black,minimum size=1mm}}
				\vertex [right =0.3cm of l1] (r1) ;
				\vertex [below =0.3cm of r1] (r2);
				\diagram* {
					(l1) -- [] (r1),
					(r1) -- [] (r2)
				};	
			\end{feynman}\end{tikzpicture}\end{gathered}
\hspace{2mm} ,
%
\hspace{2mm}
 \frac{1}{2} \hspace{2mm}
		\begin{gathered}\begin{tikzpicture} \begin{feynman}[small, baseline=g1]
					\tikzfeynmanset{every vertex={dot,black,minimum size=1mm}}
				\vertex  (l1);
				\vertex [below =0.3cm of l1] (l2);
					\tikzfeynmanset{every vertex={empty dot,black,minimum size=1mm}}
				\vertex [right =0.3cm of l1] (r1) ;
				\vertex [below =0.3cm of r1] (r2);
				\diagram* {
					(l1) -- [] (r1),
					(l1) -- [] (r2),
				};	
			\end{feynman}\end{tikzpicture}\end{gathered}$ \ , 
			\\[2mm]
&$\hspace{16.7mm} 
%
%
\frac{1}{2} \hspace{2mm}
		\begin{gathered}\begin{tikzpicture} \begin{feynman}[small, baseline=g1]
					\tikzfeynmanset{every vertex={dot,black,minimum size=1mm}}
				\vertex  (l1);
				\vertex [below =0.3cm of l1] (l2);
					\tikzfeynmanset{every vertex={empty dot,black,minimum size=1mm}}
				\vertex [right =0.3cm of l1] (r1) ;
				\vertex [below =0.3cm of r1] (r2);
				\diagram* {
					(r1) -- [] (l1),
					(r1) -- [] (l2),
				};	
			\end{feynman}\end{tikzpicture}\end{gathered}
\hspace{2mm} , \hspace{2mm} \frac{1}{4} \hspace{2mm}
			\begin{gathered}\begin{tikzpicture} \begin{feynman}[small, baseline=g1]
					\tikzfeynmanset{every vertex={dot,black,minimum size=1mm}}
				\vertex  (l1);
				\vertex [below =0.3cm of l1] (l2);
					\tikzfeynmanset{every vertex={empty dot,black,minimum size=1mm}}
				\vertex [right =0.3cm of l1] (r1) ;
				\vertex [below =0.3cm of r1] (r2);
				\diagram* {
					(l1) -- [] (l2),
					(r1) -- [] (r2),
				};	
			\end{feynman}\end{tikzpicture}\end{gathered}
\hspace{2mm} , \hspace{2mm} \frac{1}{2} \hspace{2mm}
			\begin{gathered}\begin{tikzpicture} \begin{feynman}[small, baseline=g1]
					\tikzfeynmanset{every vertex={dot,black,minimum size=1mm}}
				\vertex  (l1);
				\vertex [below =0.3cm of l1] (l2);
					\tikzfeynmanset{every vertex={empty dot,black,minimum size=1mm}}
				\vertex [right =0.3cm of l1] (r1) ;
				\vertex [below =0.3cm of r1] (r2);
				\diagram* {
					(l1) -- [] (r1),
					(l2) -- [] (r2),
				};	
			\end{feynman}\end{tikzpicture}\end{gathered}$
			\\[1mm]
$\mathcal{O}^{\scaleto{(8)}{7pt}}_{2} \hspace{2mm} = \hspace{2mm} \minus\frac{1}{2} \hspace{2mm} 
			\begin{gathered}
			\begin{tikzpicture}	
			\begin{feynman}[small, baseline=g1]
					\tikzfeynmanset{every vertex={dot,black,minimum size=1mm}}
				\vertex  (g1);
				\vertex [right =0.5cm of g1] (g2) ;
				\diagram* {
					(g1) -- [out=35,in=145] (g2) -- [out=-145,in=-35] (g1),
					(g1) -- [] (g2)
				};	
			\end{feynman}
			\end{tikzpicture}
			\end{gathered}$
&$\mathcal{O}^{\scaleto{(8)}{7pt}}_{2} \hspace{2mm} = \hspace{2mm} \minus \hspace{1mm}
			\begin{gathered}
			\begin{tikzpicture}	
			\begin{feynman}[small, baseline=g1]
					\tikzfeynmanset{every vertex={dot,black,minimum size=1mm}}
				\vertex  (g1);
					\tikzfeynmanset{every vertex={empty dot,black,minimum size=1mm}}
				\vertex [right =0.5cm of g1] (g2);
				\diagram* {
					(g1) -- [out=35,in=145] (g2) -- [out=-145,in=-35] (g1),
					(g1) -- [] (g2)
				};
			\end{feynman}
			\end{tikzpicture}
			\end{gathered}$\\[1mm]
\hline
\end{tabular}
\caption{Multigraph (off-shell) bases at mass dimension 6 and 8 used in the calculation of the ADT at mass dimension eight.}
\label{table:1}
\end{table}

\begin{table}\centering
\begin{tabular}{ |l|l| } 
\hline
	 \textbf{Real scalar theory} & \textbf{Complex scalar theory} \\
\hline
&\\[-2mm]
$\mathcal{O}^{\scaleto{(6)}{7pt}c}_{\scaleto{6}{5pt}} = \mathcal{O}^{\scaleto{(6)}{7pt}}_{\scaleto{6}{5pt}}$
&$\mathcal{O}^{\scaleto{(6)}{7pt}c}_{\scaleto{6}{5pt}} = 
	\mathcal{O}^{\scaleto{(6)}{7pt}}_{\scaleto{6}{5pt}},\quad
	\mathcal{O}^{\scaleto{(6)}{7pt}c}_4 = \minus\frac{1}{2}\mathcal{O}^{\scaleto{(6)}{7pt}}_{\scaleto{4,1}{6pt}} + \left( \mathcal{O}^{\scaleto{(6)}{7pt}}_{\scaleto{4,2}{6pt}}+\mathcal{O}^{\scaleto{(6)}{7pt}}_{\scaleto{4,3}{6pt}}\right)$\\[2mm]
$\mathcal{O}^{\scaleto{(8)}{7pt}c}_8 = \mathcal{O}^{\scaleto{(8)}{7pt}}_8$
&$\mathcal{O}^{\scaleto{(8)}{7pt}c}_8 = \mathcal{O}^{\scaleto{(8)}{7pt}}_8,\quad	
	\mathcal{O}^{\scaleto{(8)}{7pt}c}_{\scaleto{6}{5pt}} = 	\minus\frac{2}{3}\mathcal{O}^{\scaleto{(8)}{7pt}}_{\scaleto{6,1}{6pt}} + \left( 
							\mathcal{O}^{\scaleto{(8)}{7pt}}_{\scaleto{6,2}{6pt}}+\mathcal{O}^{\scaleto{(8)}{7pt}}_{\scaleto{6,3}{6pt}}\right)		$\\[3mm] 
$\mathcal{O}^{\scaleto{(8)}{7pt}c}_4 = 	\mathcal{O}^{\scaleto{(8)}{7pt}}_{\scaleto{4,1}{6pt}} -2 \mathcal{O}^{\scaleto{(8)}{7pt}}_{\scaleto{4,2}{6pt}} + 6 \mathcal{O}^{\scaleto{(8)}{7pt}}_{\scaleto{4,3}{6pt}}$
&$\mathcal{O}^{\scaleto{(8)}{7pt}c}_{4}(x,y) = 	
								(5x-y)\mathcal{O}^{\scaleto{(8)}{7pt}}_{\scaleto{4,1}{6pt}}	
								+2x\mathcal{O}^{\scaleto{(8)}{7pt}}_{\scaleto{4,2}{6pt}}		
								+2x\mathcal{O}^{\scaleto{(8)}{7pt}}_{\scaleto{4,3}{6pt}}$\\ 
&\hspace{4mm}$	-4x\mathcal{O}^{\scaleto{(8)}{7pt}}_{\scaleto{4,4}{6pt}}	
				-4x\mathcal{O}^{\scaleto{(8)}{7pt}}_{\scaleto{4,5}{6pt}}
								+(4y-16x)\mathcal{O}^{\scaleto{(8)}{7pt}}_{\scaleto{4,6}{6pt}}$\\ 

&\hspace{4mm}$+(4y-16x)\mathcal{O}^{\scaleto{(8)}{7pt}}_{\scaleto{4,7}{6pt}}
								+4y\mathcal{O}^{\scaleto{(8)}{7pt}}_{\scaleto{4,8}{6pt}}	
								+(36x-8y)\mathcal{O}^{\scaleto{(8)}{7pt}}_{\scaleto{4,9}{6pt}}$\\[2mm]
\hline
\end{tabular}
\caption{Bases of conformal primaries up to mass dimension 8. 
$\mathcal{O}^{\scaleto{(8)}{7pt}c}_{4}(x,y)$ spans two independent operators, given by different choices of $x$ and $y\in \bf{R}$.}
\label{table:2}
\end{table}

\subsection{Anomalous dimension tensors}\label{s:ADToffshell}

$M$-point  one-particle irreducible (1PI) off-shell Green's functions with $N$ insertions of higher dimension operators, 
\bea\label{eq:generalcorrelator}
	&&\Gamma_{\!\subM}[\O_{i_1},...,\O_{i_\subsubN}]=
	\Gamma_{\!\subM}(g;p_1,p_2,...,p_\subM)[\O_{i_1},...,\O_{i_\subsubN}]\\
	&&\qquad =
	\int \hspace{-1mm} \Big(\,
	\prod_{i=1}^M
	d^{\scaleto{D}{4.5pt}}\!x_i\;
	e^{ip_i\cdot x_i}\Big) 
	d^{\scaleto{D}{4.5pt}}\!y_1\, ... \,
	d^{\scaleto{D}{4.5pt}}\!y_\subsubN\;
	\langle0|\mathrm{T}\{\phi(x_1)...\phi(x_\subM)
	\, \O_{i_1}\!(y_1) ... \O_{i_\subsubN}(y_\subN)\}|0\rangle_{\text{1PI}}\nn \,, \quad \quad \phantom{.}
\eea
may diverge.
The divergences require the inclusion of counterterms in the bare Lagrangian,
\beq\label{eq:ZGO}
	\Z(\Gamma_{\!\subM}[\O_{i_1},...,\O_{i_\subsubN}])=
	\sum_{j} S_{i_1,...,i_\subsubN} \widetilde Z_{ji_1,...,i_\subsubN} \widetilde \O_j^b
	\,,
\eeq
 encoded by renormalization constant tensors $\widetilde Z_{ji_1,...,i_\subsubN}$.
We emphasize there is no sum on the $i_k$ indices in Eq.~\eqref{eq:ZGO}. The factor $S_{i_1,...,i_\subsubN}$ is a symmetry factor defined by the number of ways that $i_1,\ldots,i_N$ can be ordered. For example, $S_{i_1,...,i_\subsubN}=1$ if $i_1 = i_2 = \ldots = i_N$ and is equal to $N!$ if all $i_k$ indices are distinct. This is a choice of definition of $\widetilde Z_{ji_1,...,i_\subsubN}$ which makes subsequent equations more symmetric.
The tildes on the renormalization constant tensor 
$\widetilde Z_{ji_1,...,i_\subsubN}$ 
and operators $\widetilde \O_j^b$  in Eq.~\eqref{eq:ZGO}
reflect the fact that counterterms to off-shell correlators of physical operators generally span the {off-shell basis}. 
However, to obtain the ADT in the physical basis, we only need to consider Green's functions that include insertions of operators in a chosen physical basis (see \cite{Cao:2021cdt}), hence the absence of tildes on the operators in Eq.~\eqref{eq:generalcorrelator} and the lhs of Eq.~\eqref{eq:ZGO}.

In the physical basis, the bare couplings $c_i^{b(n)}$ are functions of the bare couplings of off-shell operators. Each of them can be expanded in terms of the renormalized couplings of the physical basis, such that 
\beq
\label{eq:Zmixing}
c_i^{b(n)}=\sum_j Z_{ij}^{(n)} c_j^{(n)}+
\sum_{\substack{n_1+n_2\\[0.5mm]=n+4}}\sum_{j,k} \, Z_{ijk}^{(n_1n_2)} c_j^{(n_1)} c_k^{(n_2)}+ O(c^3) \,.
\eeq
From this expansion follows the definition of the ADM and ADT that encode the scale dependence of the renormalized couplings $c^{(n)}_i(\mu)$:
\beq \label{eq:gammadef}
\mu\frac{d }{d\mu}c_{i}^{(n)}=\sum_j  \gamma^{(n)}_{ij}c_{j}^{(n)}
+\sum_{\substack{n_1+n_2\\[0.5mm]=n+4}}\sum_{j,k} \, \gamma^{(n_1n_2)}_{ijk}c_{j}^{(n_1)}c_{k}^{(n_2)}+O(c^3)\,.
\eeq
Requiring that the bare couplings $c_i^{b(n)}$ be independent of the renormalization scale $\mu$ one derives
 \begin{align}\label{iterative}
	 \sum_j Z_{ij}^{(n)}\ \mu\frac{d}{d\mu}c_j^{(n)}
	 & \quad = \quad - \sum_j \,
	 \mu\frac{d Z_{ij}^{(n)}}{d \mu} c_j^{(n)}
	 	 \\[1mm] &  \hspace{-13mm}
	 - \sum_{\substack{n_1+n_2\\[0.5mm]=n+4}}
	 \sum_{j,k}
	 \Bigg(\mu\frac{d Z_{ijk}^{(n_1n_2)}}{d \mu} c_j^{(n_1)}c_k^{(n_2)}
	 +Z_{ijk}^{(n_1n_2)}\mu\frac{d c_j^{(n_1)}}{d \mu} c_k^{(n_2)}
	  +Z_{ijk}^{(n_1n_2)}c_j^{(n_1)}\mu\frac{d c_k^{(n_2)}}{d \mu}\Bigg) 
	  \nn\\\nn&  \hspace{-13mm}
	  -O(c^3) \, ,
 \end{align}
which implies that
	\begin{equation}
	\gamma_{ij}^{(n)}=
	- 
	\sum_k
	\beta(g,\eps) \, 
	\big(Z^{(n)}\big)^{\text{-} \scaleto{1}{5pt}}_{ik} 
	\, \frac{d Z_{kj}^{(n)}}{d g} \,,
	\end{equation}
where
\beq
\beta(g,\eps) = \mu\,\frac{dg}{d\mu}
\eeq
is the $\beta$-function.
The solution for the ADTs of higher rank is more difficult to write down in closed form due to the appearance of lower order anomalous dimensions 
$\mu\,dc_i^{(n')}/d\mu$ (with $n'<n$). 
Therefore, one would need to solve iteratively for $\gamma_{ijk}^{(n_1n_2)}$ at fixed $n$ in terms of the anomalous dimensions at smaller mass dimension.%
	\footnote{This holds in an  EFT without relevant couplings. In contrast, for example, in an EFT with a massive particle, combinations of operators at dimension $n+a$ for nonnegative $a$ can renormalise operators of dimension $n$, proportional to $\left(\frac{m^2}{\Lambda^2}\right)^a\frac{1}{\Lambda^{n'\text{-}\,4}}$. In this case, the anomalous dimensions can be obtained order by order in $m^2/\Lambda^2$.}
The tensor for the mixing of two dimension-6 couplings into a dimension-8 coupling is given by
	\beq \label{ADtensdim8}
		\gamma_{ijk}^{(6,6)} = 
		 -\sum_l\,
		 \beta(g,\eps)\big(Z^{(8)}\big)^{\text{-} \scaleto{1}{5pt}}_{il}
			\, \frac{dZ_{ljk}^{(6,6)}}{dg}
		- \sum_{l,m}
		\big(Z^{(8)}\big)^{\text{-} \scaleto{1}{5pt}}_{il}
			\left(Z_{lmk}^{(6,6)}\gamma_{mj}^{(6)} 
					+ Z_{ljm}^{(6,6)}\gamma^{(6)}_{mk}\right)\,.
	\eeq

The calculation of the ADTs can be simplified by realizing that anomalous dimensions (and the $\beta$-function) are free of $1/\eps$ poles \cite{tHooft1973}. 
In addition, the normalization of bare operators in Eq.\ \eqref{bareoperators} ensures that 
the renormalized couplings $c_i^{(n)}$ remain dimensionless in $D=4- 2\eps$ space-time dimensions. 
This means that the anomalous dimensions also do not depend on positive powers of $\eps$. That is, $\mu \, {d c^{(n)}_i}/{d\mu}$ does not depend on $\eps$ at all. Observing in addition the relations
\begin{align}
    Z^{(n)}_{ij} &= \delta_{ij} + O(1/\eps) \, , 
    \hspace{14mm}
     \left(Z^{(n)}_{ij}\right)^{-1} = \delta_{ij} + O(1/\eps) \, , \nn\\
    Z^{(6,6)}_{ijk} &= O(1/\eps) \, ,  
    \hspace{10mm} \text{and} \hspace{10mm} 
    \beta(g,\eps) = -2 \, \eps \, g + \beta(g,0)\,,	
    \label{epsCountingZ}
\end{align}
(more generally, $Z_{ijk...}=O(1/\eps)$)
we conclude that the first sum in Eq.\ \eqref{ADtensdim8}
contains all information to determine the anomalous dimension:
\beq\label{ADdim6}
\gamma_{ijk}^{(6,6)} 
    = 
    -
    \Bigg(
    \sum_l\,
    		 \beta(g,\eps)\big(Z^{(8)}\big)^{\text{-} \scaleto{1}{5pt}}_{il}
    			\, \frac{dZ_{ljk}^{(6,6)}}{dg} 
	\Bigg) 
    			\Bigg|_{\eps^0}
    = 2g \frac{dZ^{(6,6)[1]}_{ijk}}{dg}
    \,,
\eeq
where $Z^{(6,6)[1]}_{ijk}$ is the residue of the $1/\eps$ pole of $Z^{(6,6)}_{ijk}$. The rest of the terms in Eq.\ \eqref{ADtensdim8}, namely the higher order poles in $Z^{(6,6)}_{ijk}$, the poles in 
$\big(Z^{(n)}\big)^{\text{-} \scaleto{1}{5pt}}_{ij}$, the $\beta$-function and the lower order anomalous dimensions, cancel.

By a similar argument, Eq.\ \eqref{ADdim6} generalizes to arbitrary number of EFT insertions,
\beq \label{eq:genAD}
    \gamma_{ij_1...j_\subsubN}^{(n_1...n_\subsubN)} = 2g \frac{dZ^{(n_1...n_\subsubN)[1]}_{ij_1...j_\subsubN}}{dg},
\eeq
which quantifies the mixing of $c^{(n_1)}_{j_1},...,\,c^{(n_\subsubN)}_{j_\subsubN}$ into $c^{(n)}_i$ when varying the renormalization scale. 
Although this provides a straightforward way to calculate the ADTs, it is sometimes useful to explicitly keep also the higher order terms in Eq.\ \eqref{ADtensdim8}, to confirm that they consistently cancel. 
Indeed, this was used as a consistency check on the calculation of the ADTs provided in section \ref{s:results}.

\subsection{Renormalization from amplitudes
}\label{s:ADTFF}

While our calculations are performed off shell, the proof of the non-renormalization theorem in the next section, which concerns anomalous dimensions in an on-shell basis, uses the on-shell external kinematics of amplitudes.
To control the operators under consideration, we restrict to terms with specific Wilson coefficients, instead of the full amplitudes.
We implicitly use complex momenta to define three-point amplitudes and only consider amplitudes with number of external legs $s\geq 3$. This concept has been used often in the literature to define 3-point quantities on shell, e.g.\ in the study of the 2-loop UV divergence of Einstein Gravity \cite{Goroff:1985sz}. Since the non-renormalization theorem of section \ref{s:structure} is insensitive to the precise external particle content, we label the amplitudes only by the number of external particles, $s$.

To be precise, we define  amplitudes $A_s[\O_{i_1},...,\O_{i_N}]$ as the coefficient of $c^b_{i_1},...,c^b_{i_N}$, divided by the same symmetry factor $S_{i_1,\ldots,i_N}$ appearing in Eq.~\eqref{eq:ZGO},  in the $s$-point infrared-renormalized amplitude generated by the bare EFT Lagrangian. The symmetry factor is such that in this case it allows us to write the expansion of the full amplitude
\beq
\mathcal{A}_s =\sum_{i}  c_i^b A_s[\mathcal{O}_i] +  \sum_{i,j} c_i^b   c_j^b A_s[\mathcal{O}_i,\mathcal{O}_j] +  \sum_{i,j,k}  c_i^b   c_j^b c_k^b A_s[\mathcal{O}_i,\mathcal{O}_j,\mathcal{O}_k] +  \ldots \,.
\eeq
Note that $A_s$  includes all counterterms that renormalize the fields and the couplings of the renormalizible part of the Lagrangian, as well as a multiplicative factor, (or, more precisely, an operator in color space), that cancels infrared (IR) singularities, but it has remaining UV $1/\eps$ poles. The factorisation of IR divergences in gauge theories is well established by now \cite{Becher:2009cu,Becher:2009qa}. We will use that the IR divergences do not lead to mixing of EFT operators; a similar argument was also used in Ref. \cite{Bern:2019wie} (see also \cite{Bern:2020ikv}). 

We define renormalized amplitudes $A_s^R[\O_{i_1},...,\O_{i_N}]$ as the coefficient of $c_{i_1},...,c_{i_N}$, again divided by  the  symmetry factor $S_{i_1,\ldots,i_N}$, in the $s$-point (IR-renormalized) amplitude generated by the bare EFT Lagrangian. They are thus related to the bare amplitudes via
\begin{align}
    A_s^R[\O_j]&\coloneqq\sum_i Z_{ij}A_s[\O_i]\,,\label{FFlinear}\\\label{FFdouble}
    A^R_{s}[\O_{j_1},\O_{j_2}]
	&\coloneqq	\sum_i Z_{ij_1j_2}A_{s}[\O_i]
		+ 
		\sum_{i_1,i_2}Z_{i_1j_1}Z_{i_2j_2}A_s[\O_{i_1},\O_{i_2}]\,, \quad \text{etc.}
\end{align}
The renormalized amplitudes are thus free of (UV and IR) $1/\eps$ poles.

\section{Non-renormalization theorem for multiple insertions}
\label{s:structure}

In this section, we prove the following theorem that applies to general ADTs using properties of (on-shell) Feynman diagrams.
It assumes only the minimal subtraction scheme of dimensional regularisation, the absence of relevant couplings, and an on-shell physical basis (see the definition in section \ref{s:eftlagr}). There is a more detailed remark concerning the last point in section \ref{TheoremRemarks} below.

\begin{thm}\label{thmnew}
Let 
$\O_{{i_1}}^{(n_1)},...,\O_{i_N}^{(n_N)}$ 
be $N$ operators of lengths $\ell_1,...,\ell_N$ and dimensions $n_1,...,n_N$ mixing into an operator $\O_s^{(n)}$ of length $s$ and dimension ${n=4+\sum_{i=1}^{N} (n_i - 4)}$. 
The loop order at which the corresponding anomalous dimension is first non-zero is bounded by
	\begin{equation} \label{boundconclusion}
		L \geq \text{\emph{max}}\!\left(
			1, \ 
			\frac{1}{2}\left(\sum_{i=1}^{N} \ell_i - s\right) - N + 1, \ 
			\text{\emph{max}}(\ell_1,...,\ell_N) - s + 1
			\right)\,.
	\end{equation}
\end{thm}
Restricting Theorem \ref{thmnew} to single insertions ($N=1$) results in the corollary
\begin{cor}\label{thmbern}
(ADM bound) 
Let $\O_{\ell}$ and $\O_{s}$ be operators at the same mass dimension, with lengths $\ell \geq s$.
The minimal loop order for $\O_{\ell}$ to mix into $\O_{s}$ is 
\beq 
	L \geq \ell-s+1 \,.		
\eeq 
\end{cor}
This corollary was first proven in \cite{Bern:2019wie}.
Although it follows from the more general theorem, we shall prove it separately to introduce our methodology in a simpler setting, before proving Theorem \ref{thmnew}, which is new, in full generality.

The second bound in Eq.\ \eqref{boundconclusion} follows simply from existence conditions of Feynman diagrams, as we will now show. 
All $s$-point Feynman diagrams contributing to the $S$-matrix with insertions of $\O_{i_1},...,\O_{i_N}$ satisfy
\beq\label{eq:graphcounting}
2V_R+\sum_{i=1}^N \ell_i\leq s+2I\,,
\eeq
where $I$ is the number of edges (propagators, i.e. not external legs) in the graph, and 
$V_R$ is the number of vertices of renormalizable operators, which by assumption have at least length two.
Using the textbook (e.g.~\cite{Peskin:1995ev}) relation ${L=I-V+1}$ (for connected diagrams), with the number of vertices $V = V_R+ N$, it then follows that
\beq
\label{minbound}
L\geq \frac{1}{2}\left(\sum_{i=1}^N \ell_i-s \right)-N+1
\,,
\eeq
which is the second bound in Eq.\ \eqref{boundconclusion}. 
The absence of diagrams when this bound is not satisfied implies the absence of counterterms in the coefficient of $\O_s$ proportional to $c_{i_1}...c_{i_N}$.
Note that this argument requires an on-shell physical basis, such that $\O_s$ has a non-zero (independent contact-term) contribution to the tree-level $s$-point amplitude.
Only the third bound in Theorem \ref{thmnew}, along with its corollary, remains now to be proved.

The third bound in Theorem \ref{thmnew} is non-trivial because it excludes loop orders at which non-zero (off-shell) Feynman diagrams with the external particle content of $\O_s$ and operator insertions of $\O_{i_1},...,\O_{i_N}$ do exist.
We will show that these diagrams all have a particular form which causes them to vanish on shell, and as a result they do not contribute to the mixing.
This is summarized in the following lemma:
\begin{lem}\label{lemnonlinear}
For $\text{\emph{max}}(\ell_1,...,\ell_N)>s$,
the  amplitude
with $N$ operator insertions and $s$ external legs,
 $A_s[\O_{i_1},...,\O_{i_N}]$, 
 vanishes for all loop orders 
 \beq
 L\leq \text{\emph{max}}(\ell_1,...,\ell_N)-s\,.
 \eeq
\end{lem} 
This Lemma relies on the fact that diagrams with a cut-vertex (meaning that its removal would disconnect the graph) attached to a loop-level subdiagram $\gamma_p$ with only one external momentum $p$, i.e.\ diagrams of the form
\begin{align} \label{ScalelessGraph}
    \hspace{0mm}
    \begin{gathered}\begin{tikzpicture}\begin{feynman}[small]
    \vertex at (0.5,0);
    \tikzfeynmanset{every vertex={dot,black,minimum size=1.5mm}}
    \vertex (b3) at (0,0);
       \tikzfeynmanset{every vertex={dot,minimum size=0mm}}
    \vertex at (0,-0.25) 
    ;
    \vertex at (-2.7,-0.2) {.};
    \vertex at (-2.7,0.2) {.};
    \vertex at (-2.7,0.0) {.};
    \vertex at (-2.18,0.38) (lefttop);
        \vertex at (-2.18,-0.38) (leftbottom);
    \vertex at (-3,0.5) (aaaa4) {};
    \vertex at (-3,-0.5) (aaaa5) {};
         \vertex at (-1.667,0) {$\gamma'$};
    \vertex at (-1,0) (leftl);
    \vertex at (-2.333,0) (rightl);
    \vertex at (-1.120,0.36) (upl);
    \vertex at (-1.120,-0.36) (downl);
	\vertex at (-0.78,-0.13) {.};
	\vertex at (-0.78,0.13) {.};
	\vertex at (-0.78,0.0) {.};
    \vertex at (1.120,0.36) (up);
    \vertex at (1.120,-0.36) (down);
    \vertex at (0.78,-0.13) {.};
    \vertex at (0.78,0.13) {.};
    \vertex at (0.78,0.0) {.};
    \vertex at (1,0) (left);
    \vertex at (2.333,0) (right);
     \vertex at (1.667,0) {$\gamma_{p}$};
    \vertex at (2.35,0.0) (upr);
    \vertex at (3.1,0.0) (uprr) {$p$};
    \vertex at (2.2,-0.36) (downr);
    \vertex at (2.85,-0.45) (downrr);
    %
    \diagram* {
    	(lefttop) -- [] (aaaa4),
    	(aaaa5) -- [] (leftbottom),
    	(b3) -- [] (up),
    	(b3) -- [] (down),
    	    	(b3) -- [] (upl),
    	    	(b3) -- [] (downl),
    	(left) -- [half left] (right) -- [half left] (left),
    	(leftl) -- [half left] (rightl) -- [half left] (leftl),
    	(upr) -- [] (uprr),
    };
    \end{feynman}\end{tikzpicture}\end{gathered} \,,
\end{align}
are scaleless (hence zero) for on-shell external momenta.

Let us now prove Lemma \ref{lemnonlinear}. Consider a general (potentially non-1PI) Feynman diagram $G_s$ contributing to the  amplitude $A_s[\O_{i_1},...,\O_{i_N}]$ with $s$ external legs and $N$ insertions. Assume without loss of generality  $\ell_1=\text{max}(\ell_1,...,\ell_N)$. The EFT vertex $\O_{i_1}$ is connected to $E$ external lines and by $\ell-E$ internal edges to a subgraph $\gamma$, which contains $k$ disconnected pieces with the other $N-1$ EFT vertices. 

\begin{align*}
  G_s 
  &= \ \ \text{$E$} \, {\Bigg \{}
    \hspace{-2.5mm}
    {
    \begin{gathered}\begin{tikzpicture}\begin{feynman}[small]
    \vertex at (0.5,0);
    \tikzfeynmanset{every vertex={dot,black,minimum size=1.5mm}}
    \vertex (b3) at (0,0);
       \tikzfeynmanset{every vertex={dot,minimum size=0mm}}
       \vertex at (0,0.8) {};
    \vertex at (0,-0.25) {\tiny $\O_{i_1}$};
    \vertex at (-0.72,-0.2) {.};
    \vertex at (-0.72,0.2) {.};
    \vertex at (-0.72,0.0) {.};
    \vertex at (-1,0.5) (aaaa4) {};
    \vertex at (-1,-0.5) (aaaa5) {};
    \vertex at (1.120,0.36) (up);
    \vertex at (1.120,-0.36) (down);
    \vertex at (0.78,-0.13) {.};
    \vertex at (0.78,0.13) {.};
    \vertex at (0.78,0.0) {.};
    \vertex at (1,0) (left);
    \vertex at (2.333,0) (right);
     \vertex at (1.667,0) {$\gamma$};
    \vertex at (2.2,0.36) (upr);
    \vertex at (2.85,0.45) (uprr);
    \vertex at (2.2,-0.36) (downr);
    \vertex at (2.85,-0.45) (downrr);
    \vertex at (2.67,-0.2) {.};
    \vertex at (2.67,0.2) {.};
    \vertex at (2.67,0.0) {.};
    \diagram* {
    	(b3) -- [] (aaaa4),
    	(aaaa5) -- [] (b3),
    	(b3) -- [] (up),
    	(b3) -- [] (down),
    	(left) -- [half left] (right) -- [half left] (left),
    	(upr) -- [] (uprr),
    	(downr) -- [] (downrr)
    };
    \end{feynman}\end{tikzpicture}\end{gathered}
    }
   \hspace{-0.15mm} {\Bigg \}} \, s{-}E
    \\
%
%
%
    &= \ \ \text{$E$}\,  {\Bigg \{}
\left. 
    \hspace{-3.5mm}
    {
    \begin{gathered}\begin{tikzpicture}\begin{feynman}[small]
    \vertex at (0.5,0);
    \tikzfeynmanset{every vertex={dot,black,minimum size=1.5mm}}
    \vertex (b3) at (0,0);
       \tikzfeynmanset{every vertex={dot,minimum size=0mm}}
       \vertex at (0.5,0.7) {};
    \vertex at (0,-0.25) {\tiny $\O_{i_1}$};
    \vertex at (-0.72,-0.2) {.};
    \vertex at (-0.72,0.2) {.};
    \vertex at (-0.72,0.0) {.};
    \vertex at (-1,0.5) (aaaa4) {};
    \vertex at (-1,-0.5) (aaaa5) {};
    \vertex at (1.1,0.42) (up);
    \vertex at (1.05,0.32) {\tiny .};
    \vertex at (1.2,0.28) (up2);
    \vertex at (1.33,0.45) {$\scriptstyle \gamma_{_{\scaleto{1}{3pt}}}$};
    \vertex at (1.1,0.45) (right1);
    \vertex at (1.56,0.45) (left1);
%
    \vertex at (1.06,-0.12) {.};
    \vertex at (1.06,0.12) {.};
    \vertex at (1.06,0.0) {.};
    %
    \vertex at (1.1,-0.42) (down);
    \vertex at (1.05,-0.32) {\tiny .};
    \vertex at (1.2,-0.28) (down2);
    \vertex at (1.33,-0.45) {$\scriptstyle \gamma_{{\scaleto{k}{3pt}}}$};
    \vertex at (1.1,-0.45) (right2);
    \vertex at (1.56,-0.45) (left2);
    \vertex at (1.537,0.55) (upr);
    \vertex at (2.2,0.58) (uprr);
    \vertex at (2.15,0.47) {\tiny .};
    \vertex at (1.56,0.43) (upr2);
    \vertex at (2.2,0.38) (uprr2);
    \vertex at (1.537,-0.55) (downr);
    \vertex at (2.2,-0.58) (downrr);
    \vertex at (2.15,-0.485) {\tiny .};
    \vertex at (1.56,-0.43) (downr2);
    \vertex at (2.2,-0.38) (downrr2);
    \vertex at (2.05,-0.2) {.};
    \vertex at (2.05,0.2) {.};
    \vertex at (2.05,0.0) {.};
    \diagram* {
    	(b3) -- [] (aaaa4),
    	(aaaa5) -- [] (b3),
    	(b3) -- [] (up),
    	(b3) -- [] (up2),
    	(b3) -- [] (down),
    	(b3) -- [] (down2),
     	(left1) -- [half left] (right1) -- [half left] (left1),
     	(left2) -- [half left] (right2) -- [half left] (left2),
    	(upr) -- [] (uprr),
    	(upr2) -- [] (uprr2),
    	(downr) -- [] (downrr),
    	(downr2) -- [] (downrr2),
    };
    \end{feynman}\end{tikzpicture}\end{gathered}
    }
   \hspace{0mm} 
\right\}
\, s{-}E
    \ .
\end{align*}
Let $L_i$, $V_i$ and $I_i$, for $i\in \{G_s,\gamma\}$ denote the loop number, number of vertices and number of non-external edges of the indexed graph, respectively.
Noting that $V_{G_s} = V_\gamma +1$ and $I_{G_s} = I_\gamma + \ell_1-E$, we obtain
\begin{align}
    L_{G_s} &= I_{G_s} - V_{G_s} +1 \nn\\
    &= I_\gamma - V_\gamma + \ell_1-E \nn\\
    &= L_\gamma - k + \ell_1 -E \nn\\ \label{eq:boundstep}
    &\geq  - k + \ell_1 -E 
    \,,
\end{align}
where we made use 
$L_i = I_i-V_i+k_i$ (where  $k_i$ gives the number of disconnected diagrams, $k_{G_s}=1$ and $k_{\gamma}=k$).
To obtain the last line we used $L_\gamma\geq 0$.

For $G_s$ to be non-scaleless---hence potentially non-vanishing---each of the $k$ subdiagrams $\gamma_i$ must be connected to at least two external lines.
This can only happen if $k \le (s-E)/2$; using $E<s$, this results in a bound on the contribution to the mixing
$L_{G_{s}} > \ell_1-s$, or
\begin{equation}\label{eq:linearcase}
L_{G_{s}} \geq  \ell_1-s+1 =\text{max}(\ell_1,...,\ell_N)-s+1
	\,.
\end{equation}
Therefore, there is no contribution to the amplitude $A_s[\O_{i_1},...,\O_{i_N}]$ for loop orders
\beq 
L\leq \text{max}(\ell_1,...,\ell_N)-s
\eeq
and Lemma \ref{lemnonlinear} is proved.

\subsection{Proof of Corollary \ref{thmbern}}\label{s:proofofcor}
In this section, we will first prove the result for the linear mixing (corollary \ref{thmbern}), which we subsequently generalize to multiple insertions.
We will use that the anomalous dimension $\gamma_{ij_1...j_N}^{(n_1...n_N)}$ can be determined from the residue of the $1/\varepsilon$ pole of the counterterm tensor $Z_{ij_1...j_N}^{(n_1...n_N)}$, denoted 
$Z_{ij_1...j_N}^{(n_1...n_N)[1]}$,
as illustrated in Eq.\ \eqref{eq:genAD}. 
That is, higher-order poles can be ignored for the purpose of calculating anomalous dimensions. 
In the following, the subscripts on operators refer to their lengths. Additional labels to distinguish operators of the same length are suppressed, and we remark further on this point in Section \ref{TheoremRemarks}. We also suppress the superscript for mass dimension.

For the mixing of an $\ell$-point operator $\O_\ell$ into an $s$-point operator (at the same mass dimension), consider the renormalized amplitude as written in Eq.\ \eqref{FFlinear}.
For brevity, let us define the shorthand $A_s[\O_i]\equiv A_{s,i}$ for an $s$-point  amplitude with an insertion of $\O_i$. The counterterms $Z_{ij}$ and amplitudes $A_{s,i}$ can be expanded in powers of 
$1/\varepsilon$ (see Eq.\ \eqref{epsCountingZ}),
\beq
Z_{ij}=\delta_{ij}+\sum_{n}\frac{1}{\varepsilon^n}Z_{ij}^{[n]},
\qquad 
A_{s,i}=A^{[0]}_{s,i}+
\sum_{n}\frac{1}{\varepsilon^n}A^{[n]}_{s,i}\,.
\eeq
The renormalized amplitude is free of $1/\eps$ poles.
This means that
\beq \label{epsPoleFF}
0=K_{1/\eps}\left(A^{R}_{s,\ell}\right)
=\sum_i Z_{i\ell}^{[1]}A^{[0]}_{s,i}+A^{[1]}_{s,\ell}\,,
\eeq
where the operator $K_{1/\eps}$ projects out only the residue of the $1/\eps$ pole, and the sum runs over all operators in the basis at this mass dimension.

In principle, Eq.\ \eqref{epsPoleFF} involves both UV and IR counterterms. However, we shall use this equation only at $L\leq \ell-s$ loops, such $A_{s\ell}^{(L)}$ vanishes  by Lemma \ref{lemnonlinear}. As noted in Section \ref{s:ADTFF}, this implies that IR singularities are absent up to these loop orders, since there is no lower-order amplitude for it to factorize onto.
At these loop orders, we rewrite Eq.\ \eqref{epsPoleFF} as
\beq\label{eq:Lloop0}
0=
\sum_{m=1}^{L}
\sum_{i}
Z_{i\ell}^{[1](m)}
A_{s,i}^{[0](L-m)}+\underbrace{A^{[1](L)}_{s,\ell}}_\text{\tiny{=0, Lem.\ \ref{lemnonlinear}}}\,,
\quad
\text{for $L\leq \ell-s$.}
\eeq
The sum over the operators can also be taken apart to set further terms to zero using Lemma \ref{lemnonlinear}, and we will suppress the superscript for the $1/\eps$ expansion ($[1]$ on  $Z$ and $[0]$ on $A$) from now on, as these remain the same everywhere,
\begin{eqnarray}\label{eq:Lloop}
\begin{split}
0=&
 \sum_{m=1}^{L}
  \l(\sum_{i< L-m+s}
 Z_{i\ell}^{(m)}
  A_{s,i}^{(L-m)}
  +\sum_{i \geq L-m+s}
  Z_{i\ell}^{(m)}
  \underbrace{A_{s,i}^{(L-m)}}_\text{\tiny{=0, Lem.\ \ref{lemnonlinear}}}\r)\\
  =&
  \sum_{m=1}^{L-1} \left(
  \sum_{i<L-m+s}
 Z_{i\ell}^{(m)}
  A_{s,i}^{(L-m)} \right)+
 \sum_{i<s}
  Z_{i\ell}^{(L)}
  A_{s,i}^{(0)}
  +Z_{s\ell}^{(L)}
  A_{s,s}^{(0)}\,,
 \end{split}
 \end{eqnarray}
 where in the last line we have additionally decomposed the $m=L$ term in the sum over $i$ to expose the counterterm $Z_{s\ell}^{(L)}$ of interest, multiplied by a contact term $A_{s,s}^{(0)}$. The other terms all contain $Z^{(m)}_{i\ell}$ at $m<L$ or $Z^{(L)}_{i\ell}$ with $i<s$. 

From Eq.\ \eqref{eq:Lloop}, we can prove the corollary inductively on the number of external legs $s$ and the loop level $L$. Here we instead present an linear algebraic proof, which is equivalent to the induction. Given any value of $\ell$, the number of master equations (Eq.\ \eqref{eq:Lloop}) is the same as the number of $Z_{s\ell}^{(L)}$ with indices satisfying 
$1 \leq L\leq \ell-s$ and $s\geq 3$. If we organize the target $Z_{s\ell}^{(L)}$ into a vector $x$ with the order winding in the $s-L$ space as
\beq
x=\l(Z_{3,\ell}^{(1)},...,Z_{\ell-1,\ell}^{(1)},Z_{3,\ell}^{(2)},...,Z_{\ell-2,\ell}^{(2)},...,Z_{3,\ell}^{(\ell-3)}\r)\, ,
\eeq
then the set of master equations can be written in the matrix form
\beq
M\cdot x=0\,,
\eeq
where $M$ is a square, lower triangular matrix, whose entries are amplitudes. The matrix $M$ is full rank because the diagonal terms are nonvanishing contact terms in an on-shell basis.%
\footnote{In a physical basis that is not an on-shell basis, the contact terms $F_s^{(0)}$ of operators with the same number of fields will be linearly dependent, so the proof does not apply. A more detailed remark is made at the end of the proof.}
Therefore we only have null solutions, which shows that $Z_{s\ell}^{[1](L)}=0,\forall\ L\leq \ell-s$, $L\geq 1, s\geq 3$.
Because the anomalous dimension can be determined from the $1/\eps$ pole of the counterterm, this proves Corollary \ref{thmbern}.

\subsection{Proof of Theorem \ref{thmnew}}

\paragraph{Two insertions.}

We now prove the third bound of Theorem \ref{thmnew} for the case of $N=2$; the second bound has already been proven above. The logic of the proof can easily be generalized to arbitrary $N$. Let us define the shorthand $A_s[\O_{i_1},...,\O_{i_N}]\equiv A_{s,i_1...i_N}$ for an $s$-point amplitude with $N$ insertions of $\O_{i_1},...,\O_{i_N}$. 
The renormalized amplitude with insertions of $\O_{\ell_1},\O_{\ell_2}$ can be expanded as in Eq.\ \eqref{FFdouble}.
Noting that $Z_{i\ell_1\ell_2}=O(1/\eps)$ (see Eq.\ \eqref{epsCountingZ}), the contributions to the zero residue of the $1/\eps$ pole are
\begin{equation}\label{2operators1/eResidue}
 0=K_{1/\varepsilon}(A^R_{s,\ell_1\ell_2}) =
  A_{s,\ell_1\ell_2}^{[1]}
 +\sum_i \left(Z_{i\ell_1}^{[1]}A_{s,i\ell_2}^{[0]}
 +Z_{i\ell_2}^{[1]}A_{s,i\ell_1}^{[0]}
 +Z_{i\ell_1\ell_2}^{[1]}A_{s,i}^{[0]} \right)\,.
\end{equation}
For concreteness, but without loss of generality, let us assume that $\ell_1\geq (\ell_2,s)$. 
We only use Eq.\ \eqref{2operators1/eResidue} for loop orders $L\leq \ell_1-s$.
At these orders, $A_{s,\ell_1\ell_2}^{[1](L)}$ and $A_{s,i\ell_1}^{[0](L)}$ vanish by Lemma \ref{lemnonlinear}.
Suppressing again the superscript for the $1/\eps$ expansion ($[1]$ on  $Z$ and $[0]$ on $A$), this gives
\begin{equation} \label{2operators1/eResidue-2}
 0 =
 \sum_{m=1}^{L}
 \sum_i \left(Z_{i\ell_1}^{(m)}A_{s,i\ell_2}^{(L-m)}
 +Z_{i\ell_1\ell_2}^{(m)}A_{s,i}^{(L-m)} \right)\,.
\end{equation}
Finally, we note that $A_{s,i}^{(L-m)}=0=A_{s,i\ell_2}^{(L-m)}$ for $L-m+s\leq i$ if $L-m>0$, and 
$A_{s,i}^{(0)}=0=A_{s,i\ell_2}^{(0)}$ for $i>s$ by Lemma  \ref{lemnonlinear}.
Therefore, Eq.\ \eqref{2operators1/eResidue-2} rewrites into
\begin{align}\label{2operators1/eResidue-3}
 0 =&
 \sum_{i\leq s} Z_{i\ell_1}^{(L)}A_{s,i\ell_2}^{(0)}
  +\sum_{i<s}Z_{i\ell_1\ell_2}^{(L)}A_{s,i}^{(0)} 
  +Z_{s\ell_1\ell_2}^{(L)}A_{s,s}^{(0)} 
  +
  \nn\\
  &+
 \sum_{m=1}^{L-1}
 \sum_{i<L-m+s} \left(Z_{i\ell_1}^{(m)}A_{s,i\ell_2}^{(L-m)}
 +Z_{i\ell_1\ell_2}^{(m)}A_{s,i}^{(L-m)} \right)\,.
\end{align}

Eq.\ \eqref{2operators1/eResidue-3} contains terms with $Z_{i\ell_1}^{(m)}$ 
which are all zero by the proof in the previous subsection. More generally, counterterms with fewer insertions can be argued to be zero by induction on the number of insertions. 
Exactly analogous to the argument after Eq.\ \eqref{eq:Lloop}, it then follows that 
$Z_{s\ell_1\ell_2}^{(L)}=0$ for $s<\ell_1$ and $L\leq \ell_1-s=\text{max}(\ell_1,\ell_2)-s$.

\paragraph{Beyond the bounds.}
It is instructive to identify the form of non-scaleless diagrams that minimally satisfy the constraint at $N=2$, depending on the value of $s$ relative to $\ell_1$ and $\ell_2$, assuming $\ell_1\geq\ell_2$. 
The graphs we draw assume that there are only 4-point interactions in the renormalizable Lagrangian, but this is easily generalized to three-point interactions. It is possible in specific theories that the following graphs cannot be drawn with particular particle content. In this theory-specific case, the bounds can be further improved.

When $s\geq \ell_1+\ell_2-4$, the diagram
 \beq
\quad
 G \, = \,
     \begin{gathered}\begin{tikzpicture}\begin{feynman}[small]
 	    \vertex at (0,0.58) {};
 	    \tikzfeynmanset{every vertex={dot,black,minimum size=1.5mm}}
 	    \vertex (EFTmax) at (0,0);
 	    \vertex at (1.5,0) (EFT2);
 	       \tikzfeynmanset{every vertex={dot,minimum size=0mm}}
 		\vertex at (-0.4,0.17) (intop);
 		\vertex at (-0.4,-0.17) (inbottom);
 		\vertex at (1.9,0.17) (outtop);
 		\vertex at (1.9,-0.17) (outbottom);
 		\vertex at (-0.33,0.065)  {\tiny $.$};
 		\vertex at (-0.33,0) 	  {\tiny $.$};
 		\vertex at (-0.33,-0.065) {\tiny $.$};
 		\vertex at (1.2,0) (right);
 		\vertex at (0.7,0) (left);
 	    \vertex at (1.83,0.065)  {\tiny $.$};
 	    \vertex at (1.83,0) 	 {\tiny $.$};	    	   
  	    \vertex at (1.83,-0.065) {\tiny $.$};
 		\vertex at (0.1,-0.32) {\tiny $\O_{i_1}$};
 		\vertex at (1.68,-0.32) {\tiny $\O_{i_2}$};
 		\vertex at (0.95,0.27) (phi41);
 		\vertex at (0.87,0.42) (phi41in);
 		\vertex at (1.03,0.42) (phi41out);
 		\vertex at (0.55,0.27) (phi42);
 		\vertex at (0.47,0.42) (phi42in);
 		\vertex at (0.63,0.42) (phi42out);
  	    \vertex at (0.75,0.35)  {\tiny $.$};
  	    \vertex at (0.82,0.35) 	 {\tiny $.$};	    	   
   	    \vertex at (0.68,0.35) {\tiny $.$};
 	    \diagram* {
 	    	(phi41in) --[] (phi41) --[] (phi41out),
 	    	(phi42in) --[] (phi42) --[] (phi42out),
 			(EFTmax) --[out=-38,in=-142] (EFT2),
 			(EFTmax) --[out=38,in=142] (EFT2),
 			(outtop) --[] (EFT2) --[] (outbottom),			
 			(intop) --[] (EFTmax) --[] (inbottom),
 	    };
 	    \end{feynman}\end{tikzpicture}\end{gathered}\quad ,
 \eeq
 
 \vspace{-3mm}\noindent
contributes at 1-loop. When the second bound of Eq.\ \eqref{boundconclusion} dominates instead, the diagram
 \beq
 \quad 
 G \, = \,
     \begin{gathered}\begin{tikzpicture}\begin{feynman}[small]
 	    \vertex at (0,0.58) {};
 	    \tikzfeynmanset{every vertex={dot,black,minimum size=1.5mm}}
 	    \vertex (EFTmax) at (0,0);
 	    \vertex at (1.5,0) (EFT2);
 	       \tikzfeynmanset{every vertex={dot,minimum size=0mm}}
 		\vertex at (-0.4,0.17) (intop);
 		\vertex at (-0.4,-0.17) (inbottom);
 		\vertex at (1.9,0.17) (outtop);
 		\vertex at (1.9,-0.17) (outbottom);
 		\vertex at (-0.33,0.065)  {\tiny $.$};
 		\vertex at (-0.33,0) 	  {\tiny $.$};
 		\vertex at (-0.33,-0.065) {\tiny $.$};
 		\vertex at (1.2,0) (right);
 		\vertex at (0.7,0) (left);
 	    \vertex at (1.83,0.065)  {\tiny $.$};
 	    \vertex at (1.83,0) 	 {\tiny $.$};	    	   
  	    \vertex at (1.83,-0.065) {\tiny $.$};
 		\vertex at (0.1,-0.32) {\tiny $\O_{i_1}$};
 		\vertex at (1.68,-0.32) {\tiny $\O_{i_2}$};
 		\vertex at (0.95,0.27) (phi41);
 		\vertex at (0.87,0.42) (phi41in);
 		\vertex at (1.03,0.42) (phi41out);
 		\vertex at (0.55,0.27) (phi42);
 		\vertex at (0.47,0.42) (phi42in);
 		\vertex at (0.63,0.42) (phi42out);
  	    \vertex at (0.75,0)  	 { $.$};
  	    \vertex at (0.75,0.12) 	 { $.$};	    	   
   	    \vertex at (0.75,-0.12)  { $.$};
 	    \diagram* {
 			(EFTmax) --[out=-38,in=-142] (EFT2),
 			(EFTmax) --[out=38,in=142] (EFT2),
 			(outtop) --[] (EFT2) --[] (outbottom),			
 			(intop) --[] (EFTmax) --[] (inbottom),
 	    };
 	    \end{feynman}\end{tikzpicture}\end{gathered}\quad ,
 \eeq
contributes at $\frac{1}{2}(\ell_1+\ell_2-s)-1$ loops.
Finally, when the third bound is strongest, a diagram that contributes has the form, for example,
\beq \label{bound3N=2}
\quad 
G \, = \,
{\scriptstyle s -2} \{  \hspace{-1.3mm}
    \begin{gathered}\begin{tikzpicture}\begin{feynman}[small]
	    \vertex at (0,0.58) {};
	    \tikzfeynmanset{every vertex={dot,black,minimum size=1.5mm}}
	    \vertex (EFTmax) at (0,0);
	    \vertex at (1.5,0) (EFT2);
	       \tikzfeynmanset{every vertex={dot,minimum size=0mm}}
	    \vertex at (0.75,-0.12) {\tiny $.$};
	    \vertex at (0.75,-0.18) {\tiny $.$};	    	   
 	    \vertex at (0.75,-0.24) {\tiny $.$};
		\vertex at (-0.4,0.17) (intop);
		\vertex at (-0.4,-0.17) (inbottom);
 		\vertex at (1.9,0.17) (outtop);
 		\vertex at (1.9,-0.17) (outbottom);
		\vertex at (-0.33,0.065) {\tiny $.$};
		\vertex at (-0.33,0) {\tiny $.$};
		\vertex at (-0.33,-0.065) {\tiny $.$};
		\vertex at (1.2,0) (right);
		\vertex at (0.7,0) (left);
	    \vertex at (0.79,0.06) {\tiny $.$};
	    \vertex at (0.88,0.06) {\tiny $.$};	    	   
 	    \vertex at (0.97,0.06) {\tiny $.$};
		\vertex at (0.1,-0.32) {\tiny $\O_{i_1}$};
		\vertex at (1.68,-0.32) {\tiny $\O_{i_2}$};
	    \diagram* {
	    	(EFTmax) --[] (EFT2),
			(EFTmax) --[out=-38,in=-142] (EFT2),
			(outtop) --[] (EFT2) --[] (outbottom),			
			(intop) --[] (EFTmax) --[] (inbottom),
			(EFTmax) --[out=70, in=110] (right) --[out=130,in=50] (EFTmax),
			(EFTmax) --[out=38,in=142] (left) --[out=160,in=20] (EFTmax)
	    };
	    \end{feynman}\end{tikzpicture}\end{gathered}\quad ,
\eeq
with $\frac{1}{2}(\ell_1-\ell_2-s+4)$  four-point vertices. 
This graph has $\ell_1-s+1$ loops.

\paragraph{General number of insertions.}
For mixing of $N$ operators $\O_{j_1},...,\O_{j_N}$ into an $s$-point operator, the proof that $Z_{sj_1...j_N}^{[1](L)}=0$ for $L\leq j_1-s$ is logically the same as the case of $N=2$.
For simplicity, we use notation $J=\{j_1,..,j_N\}$ and take $j_1=\text{max}(J)\equiv\text{max}(j_1,..,j_N)>s$. The corresponding renormalized amplitude is, schematically,
\beq \label{F^Rnonlinear}
A^R_{s,J} =
\sum_{\text{all partitions of $J$}} 
\alpha \, \l(\prod_{k=1}^n Z_{i_kJ_{k}}\r) \, A_{s,i_1...i_n}\,,
\eeq
where $\{J_a\}|_{a=1,...,n}$ is a partition of set $J$, i.e.\ $J_a \subset J$, with 
$\cup_{a=1}^{n} J_a = J$ and 
$J_a \cap J_b = \emptyset$,
and the sum runs over all possible different partitions of $J$. 
We have symmetries between indices in $Z_{i...j_l...j_m...}= Z_{i...j_m...j_l...}$ and 
$A_{s,...i_l...i_m...}=A_{s,...i_m...i_l...} $.  
The second symmetry allows us to set an order in the partition of $J$ such as $|J_1|\geq|J_2|\geq...\geq |J_n|$, where $|J_i|$ is the number of elements in set $J_i$. 
We have implicitly summed over repeated indices, and we include a potentially different numerical factor, $\alpha$, for each term. 
%
%
%
%
%
%
For example, Eq.\ \eqref{F^Rnonlinear} for three insertions is
\begin{align} 
	A^R_{s,j_1j_2j_3} 
	&=
	Z_{i,j_1j_2j_3}A_{s,i}
	+ 
	\frac{2}{3}
		(Z_{i_1j_1}Z_{i_2j_2j_3}+
		Z_{i_1j_2}Z_{i_2j_3j_1}+
		Z_{i_1j_3}Z_{i_2j_1j_2})A_{s,i_1i_2} \nn\\
	& \quad +
	Z_{i_1 j_1}Z_{i_2 j_2} Z_{i_3 j_3} A_{s,i_1i_2i_3}
\end{align}
The numerical factors will not be important for the following argument, so we will not explicitly write them in the following proof. 

The fact that the ADTs only depend on the residue of the $1/\eps$ pole helps identifying the relevant terms in Eq.\ \eqref{F^Rnonlinear}. 
$Z_{ijk...}=O(1/\eps)$ implies that products of renormalization tensors with more than two indices contain only higher poles. Only partitions with $|J_1|\geq1, |J_2|=...=|J_n|=1$ contribute to the ADTs. Eq.\ \eqref{F^Rnonlinear} can thus be rearranged as
\begin{eqnarray}
\begin{split}
A^R_{s,J}=
&\sum_{\substack{\text{all partitions of } J\\ |J_{k>1}|=1}}\l(\prod_{k=1}^n Z_{i_kJ_{k}}\r)A_{s,i_1...i_n}+(\text{terms that only include higher poles}) \, .
\end{split}
\end{eqnarray}
Considering that the residue of the $1/\eps$ pole is zero, we find
\begin{align}
 0=&K_{1/\varepsilon}(A^R_{s,J})
\nn\\
 =&A_{s,J}^{[1]}+\sum_{j_1\in J_1} Z_{iJ_1}^{[1]}A_{s,i(J-J_1)}^{[0]} + \sum_{j_1\notin J_1} Z_{iJ_1}^{[1]}A_{s,i(J-J_1)}^{[0]}\,.
\end{align}
The sum in the last line is over all possible partitions of $J$ with $|J_{k>1}|=1$, and we isolated the terms with $j_1\in J_1$ in a separate sum.
The structure of the above equation is similar to that in the last subsection. 
We will use this equation at fixed loop order up to $\text{{max}}(J)-s=j_1-s$ loops, for which $A_{s,J}^{[1]}$ and $A_{s,i(J-J_1)}^{[0]}$ (if $j_1\notin J_1$) vanish by Lemma \ref{lemnonlinear}.
Suppressing the superscript for the $1/\eps$ expansion ($[1]$ on  $Z$ and $[0]$ on $A$), this gives
\begin{equation} \label{noperators1/eResidue-2}
 0 =
 \sum_{m=1}^{L}
 \sum_i \left(\sum_{j_1\in J_1,J_1\neq J} Z_{iJ_1}^{(m)}A_{s,i(J-J_1)}^{(L-m)} 
 +Z_{iJ}^{(m)}A_{s,i}^{(L-m)} \right)\,\,,
\end{equation}
which resembles \eqref{2operators1/eResidue-2} if we replace $\ell_1$ by $J_1$ and ${\ell_1,\ell_2}$ by $J$. 
Using a similar argument, we note that $A_{s,i}^{(L-m)}=0=A_{s,i(J-J_1)}^{(L-m)} $ for $L-m+s\leq i$ if $L-m>0$, and 
$A_{s,i}^{(0)}=0=A_{s,i(J-J_1)}^{(0)} $ for $i>s$ by Lemma  \ref{lemnonlinear}.
Therefore, Eq.\ \eqref{noperators1/eResidue-2} can be rewritten as
\begin{align}\label{noperators1/eResidue-3-OLD}
 0 =&
\sum_{j_1\in J_1,J_1\neq J} \left(\sum_{m=1}^{L-1}
 \sum_{i<L-m+s} Z_{iJ_1}^{(m)}A_{s,i(J-J_1)}^{(L-m)} +\sum_{i\leq s} Z_{iJ_1}^{(L)}A_{s,i(J-J_1)}^{(0)}\right)\nn\\
 &+
 \sum_{m=1}^{L-1}
 \sum_{i<L-m+s} \left(Z_{iJ}^{(m)}A_{s,i}^{(L-m)} \right)+\sum_{i<s}Z_{iJ}^{(L)}A_{s,i}^{(0)} 
  +Z_{sJ}^{(L)}A_{s,s}^{(0)} 
  \,,
\end{align}
where we split the sum into the part containing ADTs with number of insertions less than $N$ in the first line and the part containing ADTs with $N$ insertions in the second line.

The first line of Eq.\ \eqref{noperators1/eResidue-3-OLD} contains terms with $Z_{iJ_1}^{(m)}$ 
which are all zero by induction on the number of insertions. With an argument the same as that given after Eq.\ \eqref{eq:Lloop}, it then follows that 
$Z_{sJ}^{(L)}=0$ for $s<\text{max}(J)$ and $L\leq \text{max}(J)-s$.

\subsection{Remarks on the theorem} \label{TheoremRemarks}

\paragraph{Multiple operators of the same length.}
To simplify the notation, we have suppressed indices that label different operators with the same number of fields.
When there are multiple operators with the same number of fields at the same mass dimension, i.e. 
${\mathcal{L} \supset \sum_i c_{(s,i)}\O_{(s,i)}}$, the above proves that $\sum_i Z_{(s,i)J}^{(L)[1]}A^{(0)}_{s,(s,i)}=0$. In an on-shell basis, the contact terms are linearly independent (and non-zero). This implies that each counterterm itself vanishes, $Z_{(s,i)J}^{(L)[1]}=0, \forall i$. This argument fails to hold in operators bases other than on-shell bases, when there are (linear combinations of) operators which vanish as on-shell contact terms, see below. 

Similarly, we suppressed reference to the specific external field content of amplitudes, labelling them only by the number of fields. 
In an explicit proof, one would use the fact that $A_{s,s}^{(0)}$ is only non-zero if the external particle content matches that of the inserted operator.

\paragraph{Choice of basis.}

As stated before, Theorem \ref{thmnew} only holds for on-shell bases. We stress that physical bases containing, for example, operators proportional to $\partial^2\phi$ are perfectly possible, even though they vanish as on-shell contact terms.\footnote{This was not explicitly stated in Ref.\ \cite{Bern:2019wie} where the on-shell analysis restricts to so-called minimal form factors (i.e.\ contact terms). These may result in relations $0\,\gamma = 0$, for bases with operators that vanish as on-shell contact terms. Therefore, in such cases, one requires non-minimal form factors to find that the anomalous dimension $\gamma$ potentially does not vanish.}

For example, at mass dimension six for the complex scalar, an on-shell basis consists of a 6-point operator and a 4-point operator. However, the 6-point operator can be replaced by a 2-point operator using a field redefinition,
\begin{align*}
\frac{g^2}{4} \hspace{3mm}
 \begin{gathered}
			\begin{tikzpicture}	
			\begin{feynman}[small, baseline=g1]
					\tikzfeynmanset{every vertex={dot,black,minimum size=1mm}}
				\vertex  (g1);
				\vertex [below =0.3cm of g1] (g2) ;
				\vertex [below =0.3cm of g2] (g3);
					\tikzfeynmanset{every vertex={empty dot,black,minimum size=1mm}}
				\vertex [right =0.3cm of g1] (g6);
				\vertex [below =0.3cm of g6] (g7);
				\vertex [below =0.3cm of g7] (g8);
				\diagram* {
				};	
			\end{feynman}
			\end{tikzpicture}
			\end{gathered}
 \hspace{3mm} \stackrel{\textsc{fr}}{\rightarrow}   \hspace{3mm}  
\begin{gathered}
	\begin{tikzpicture}
	\begin{feynman}[small, baseline=g4]
	\tikzfeynmanset{every vertex={dot,black,minimum size=1mm}}
	\vertex  (g1);
	\tikzfeynmanset{every vertex={empty dot,black,minimum size=1mm}}
	\vertex [right =0.3cm of g1] (g2) ;
	\tikzfeynmanset{every vertex={dot,red,minimum size=0mm}}
	\vertex [above =2.5mm of g1] (loop1);
	\vertex [above =2.5mm of g2] (loop2);
	\diagram* {
			(g1) -- [out=45,in=0] (loop1) -- [out=180,in=135] (g1),
		(g2) -- [out=45,in=0] (loop2) -- [out=180,in=135] (g2)
	};
	\end{feynman}
	\end{tikzpicture}
\end{gathered}
\hspace{2mm}
 \,,
\end{align*}
which does not affect the ADM. Therefore, in the obtained basis, the mixing of a 4-point operator into a 2-point operator already occurs at two loops for the conformal 4-point operator, and even at one loop for a generic choice of 4-point operator (see \cite{Cao:2021cdt}). This clearly violates a naive application of the theorem that would have dictated zeros up to $L\ge 3$ loops.

\section{Results: ADTs up to three loops in scalar EFT}
\label{s:results}

To exemplify the non-renormalization theorems, we have computed the ADTs at mass dimension eight for the real and the complex scalar. For this purpose, we employed the  $R^*$ method, as described further in \cite{Herzog:2017bjx,deVries:2019nsu,Beekveldt:2020kzk}, to compute the renormalisation constants in an off-shell basis---specifically the multigraph basis defined in table \ref{table:1}. The conversion into the physical basis, from which the physical ADT was extracted, was then performed using explicit field redefinitions.
The calculations were performed in $D=4-2\epsilon$ space-time dimensions using the FORCER program \cite{Ruijl:2017cxj} in FORM \cite{Vermaseren:2000nd,Ruijl:2017dtg,Ruijl:2017cxj,Ueda:2020wqk} and a private implementation of the $R^*$ operation in Maple \cite{Maple}. The timing for the most demanding of these calculations was of the order of hours on a modern 64 core machine. One could have thus easily pushed the scope of these calculations to higher loops (4 or potentially even 5) and/or mass dimensions, if there would be the interest to do so, using this setup.

Our results for the ADTs complement the results of ADMs at linear order of \cite{Cao:2021cdt}, to which we also refer for further details on the calculation.

\subsection{Real scalar mass dimension eight}

The ADT for the mixing of two insertions of dimension six operators into the dimension eight operators up to three loops is given by
$$
\gamma^{(6,6)} = 
\begin{pmatrix}
-35g+\frac{1568}{3}g^2-\left(3360\zeta_3+\frac{201425}{24}\right)g^3\\[2mm]
\textcolor{black}{0g+0g^2}-\frac{1}{432}g^3
\end{pmatrix}  .
$$
This tensor multiplies two factors of the dimension-6 coupling of $g^2\,\O_6^{(6)c}$, to give the $\mu$-dependence of the dimension-8 couplings in the basis of conformal primaries, $g^3\,\O_8^{(8)c}$ (first component) and $g\,\O_4^{(8)c}$ (second component).
Explicitly, the operators are defined in table \ref{table:2}.

The zeros in this tensor are explained by the absence of 1- and 2-loop diagrams with two insertions of $g^2\,\O^{(6)c}_6$ and the external particle content of $g\,\O^{(8)c}_4$.

\subsection{Complex scalar mass dimension eight}

The ADT at mass dimension eight in the basis of conformal primaries at both dimension 6 and 8 up to three loops is given by
\begin{equation}
\gamma^{(6,6)}_{ijk} =
	\begin{pmatrix}
		\begin{pmatrix}
			-52g+\frac{6024}{5}g^2 
			-\left(11712 \zeta_3+\frac{59783}{2}\right)g^3\ \ \ \ 
				&\frac{24}{5}g-\frac{942}{5}g^2
			+\left( \frac{12672}{5}\zeta_3+\frac{108343}{20} \right)g^3
				\\[2mm]
			\frac{24}{5}g-\frac{942}{5}g^2
		+\left( \frac{12672}{5}\zeta_3+\frac{108343}{20} \right)g^3
			 \ \ \ \ 
				&\frac{216}{5}g -\frac{1791}{10}g^2
		+\left(-\frac{10152}{5}\zeta_3+\frac{465073}{160}\right)g^3
		\end{pmatrix}_{\!\!jk}\\[8mm]
			\begin{pmatrix}
		0g+\frac{1}{5}g^2 
		-\frac{107}{60}g^3
		\ \ \ \ 
		& -\frac{9}{5}g+\frac{273}{20}g^2
		-\left( \frac{162}{5}\zeta_3+\frac{93443}{480} \right) g^3
		\\[2mm]
		-\frac{9}{5}g+\frac{273}{20}g^2
		-\left( \frac{162}{5}\zeta_3+\frac{93443}{480} \right) g^3
		 \ \ \ \ 
		&-\frac{111}{20}g + \frac{9827}{120}g^2
		-\left( \frac{6927}{20}\zeta_3+\frac{305423}{384}\right)g^3
		\end{pmatrix}_{\!\!jk}\\[8mm]
			\begin{pmatrix}
		0g+0g^2-\frac{1}{180}g^3 \ \ \ \ 
		& 	{0g+0g^2-\frac{1}{180}g^3
		}
		\\[2mm]
			{0g+0g^2-\frac{1}{180}g^3
			} \ \ \ \ 
		&-\frac{7}{20}g+\frac{17}{144}g^2
		-\left(\frac{1}{4}\zeta_3-\frac{73}{5760}\right)g^3
		\end{pmatrix}_{\!\!jk}\\[8mm]
		\begin{pmatrix}
			{0g+0g^2-\frac{2}{135}g^3} \ \ \ \ 
		& 	{0g+0g^2-\frac{11}{180}g^3
		}
		\\[2mm]
			{0g+0g^2-\frac{11}{180}g^3
			} \ \ \ \ 
		&-\frac{27}{20}g+\frac{43}{144}g^2
		-\left( \frac{3}{4}\zeta_3-\frac{35987}{51840} \right) g^3
		\end{pmatrix}_{\!\!jk}
		
	\end{pmatrix}_i \, ,
\end{equation}
which multiplies two dimension-six couplings, $c^{(6)}_j \, c^{(6)}_k$,
with the $i$ index of $c^{(6)}_i$ corresponding to the $i$th operator in the set
$$
\left\{  g^2 \, \O^{(6)c}_6 \ , \ g \, \O^{(6)c}_4      \right\},
$$
and as such contribute to the scale dependence of $c^{(8)}_i$, 
with the $i$ index of $c^{(8)}_i$ corresponding to the $i$th operator in the set
$$
\left\{
g^3 \, \O^{(8)c}_8 \ , \ 
g^2 \, \O^{(8)c}_6 \ , \ 
g \, \O^{(8)c}_4{{(1,0)}} \ , \ 
g \, \O^{(8)c}_4{{(0,1)}} \right\}.
$$
Explicitly, the operators in the chosen basis are given in table \ref{table:2}.
The two independent choices of $\O_4^{(8)c}(x,y)$ are
\begin{align*}			
%
	\mathcal{O}^{\scaleto{(8)c}{7pt}}_{4}(1,0) \hspace{2mm} &= 
	\hspace{3mm} 5 \hspace{2mm}
		\begin{gathered}
			\begin{tikzpicture}	
			\begin{feynman}[small, baseline=g1]
					\tikzfeynmanset{every vertex={dot,black,minimum size=1mm}}
				\vertex  (l1);
				\vertex [below =0.3cm of l1] (l2);
					\tikzfeynmanset{every vertex={empty dot,black,minimum size=1mm}}
				\vertex [right =0.3cm of l1] (r1) ;
				\vertex [below =0.3cm of r1] (r2);
				\diagram* {
					(l1) -- [out=30,in=150] (r1) -- [out=-150,in=-30] (l1)
				};	
			\end{feynman}
			\end{tikzpicture}
			\end{gathered}
\hspace{2mm} + \frac{1}{2} \hspace{2mm}
		\begin{gathered}
			\begin{tikzpicture}	
			\begin{feynman}[small, baseline=g1]
					\tikzfeynmanset{every vertex={dot,black,minimum size=1mm}}
				\vertex  (l1);
				\vertex [below =0.3cm of l1] (l2);
					\tikzfeynmanset{every vertex={empty dot,black,minimum size=1mm}}
				\vertex [right =0.3cm of l1] (r1) ;
				\vertex [below =0.3cm of r1] (r2);
				\diagram* {
					(l1) -- [out=-60,in=60] (l2) -- [out=120,in=-120] (l1)
				};	
			\end{feynman}
			\end{tikzpicture}
			\end{gathered}
\hspace{2mm} + \frac{1}{2} \hspace{2mm}
		\begin{gathered}\begin{tikzpicture} \begin{feynman}[small, baseline=g1]
					\tikzfeynmanset{every vertex={dot,black,minimum size=1mm}}
				\vertex  (l1);
				\vertex [below =0.3cm of l1] (l2);
					\tikzfeynmanset{every vertex={empty dot,black,minimum size=1mm}}
				\vertex [right =0.3cm of l1] (r1) ;
				\vertex [below =0.3cm of r1] (r2);
				\diagram* {
					(r1) -- [out=-60,in=60] (r2) -- [out=120,in=-120] (r1)
				};	
			\end{feynman}\end{tikzpicture}\end{gathered}
%
%
%
-4 \hspace{2mm}
		\begin{gathered}
			\begin{tikzpicture}	
			\begin{feynman}[small, baseline=g1]
					\tikzfeynmanset{every vertex={dot,black,minimum size=1mm}}
				\vertex  (l1);
				\vertex [below =0.3cm of l1] (l2);
					\tikzfeynmanset{every vertex={empty dot,black,minimum size=1mm}}
				\vertex [right =0.3cm of l1] (r1) ;
				\vertex [below =0.3cm of r1] (r2);
				\diagram* {
					(l1) -- [] (r1),
					(l1) -- [] (l2)
				};	
			\end{feynman}
			\end{tikzpicture}
			\end{gathered}
%
%
%
\hspace{2mm} -4 \hspace{2mm}
		\begin{gathered}\begin{tikzpicture} \begin{feynman}[small, baseline=g1]
					\tikzfeynmanset{every vertex={dot,black,minimum size=1mm}}
				\vertex  (l1);
				\vertex [below =0.3cm of l1] (l2);
					\tikzfeynmanset{every vertex={empty dot,black,minimum size=1mm}}
				\vertex [right =0.3cm of l1] (r1) ;
				\vertex [below =0.3cm of r1] (r2);
				\diagram* {
					(l1) -- [] (r1),
					(r1) -- [] (r2)
				};	
			\end{feynman}\end{tikzpicture}\end{gathered}
\hspace{2mm} 
%
-8 \hspace{2mm}
		\begin{gathered}\begin{tikzpicture} \begin{feynman}[small, baseline=g1]
					\tikzfeynmanset{every vertex={dot,black,minimum size=1mm}}
				\vertex  (l1);
				\vertex [below =0.3cm of l1] (l2);
					\tikzfeynmanset{every vertex={empty dot,black,minimum size=1mm}}
				\vertex [right =0.3cm of l1] (r1) ;
				\vertex [below =0.3cm of r1] (r2);
				\diagram* {
					(l1) -- [] (r1),
					(l1) -- [] (r2),
				};	
			\end{feynman}\end{tikzpicture}\end{gathered}
%
%
%
%
%
%
-8 \hspace{2mm}
		\begin{gathered}\begin{tikzpicture} \begin{feynman}[small, baseline=g1]
					\tikzfeynmanset{every vertex={dot,black,minimum size=1mm}}
				\vertex  (l1);
				\vertex [below =0.3cm of l1] (l2);
					\tikzfeynmanset{every vertex={empty dot,black,minimum size=1mm}}
				\vertex [right =0.3cm of l1] (r1) ;
				\vertex [below =0.3cm of r1] (r2);
				\diagram* {
					(r1) -- [] (l1),
					(r1) -- [] (l2),
				};	
			\end{feynman}\end{tikzpicture}\end{gathered}
%
%
%
\hspace{2mm} +18 \hspace{2mm}
			\begin{gathered}\begin{tikzpicture} \begin{feynman}[small, baseline=g1]
					\tikzfeynmanset{every vertex={dot,black,minimum size=1mm}}
				\vertex  (l1);
				\vertex [below =0.3cm of l1] (l2);
					\tikzfeynmanset{every vertex={empty dot,black,minimum size=1mm}}
				\vertex [right =0.3cm of l1] (r1) ;
				\vertex [below =0.3cm of r1] (r2);
				\diagram* {
					(l1) -- [] (r1),
					(l2) -- [] (r2),
				};	
			\end{feynman}\end{tikzpicture}\end{gathered} 
			\hspace{2mm},\\[1.5mm]
%
%
%
%
%
	\mathcal{O}^{\scaleto{(8)c}{7pt}}_{4}(0,1) \hspace{2mm} &= \hspace{3mm} 
	- \,
		\begin{gathered}
			\begin{tikzpicture}	
			\begin{feynman}[small, baseline=g1]
					\tikzfeynmanset{every vertex={dot,black,minimum size=1mm}}
				\vertex  (l1);
				\vertex [below =0.3cm of l1] (l2);
					\tikzfeynmanset{every vertex={empty dot,black,minimum size=1mm}}
				\vertex [right =0.3cm of l1] (r1) ;
				\vertex [below =0.3cm of r1] (r2);
				\diagram* {
					(l1) -- [out=30,in=150] (r1) -- [out=-150,in=-30] (l1)
				};	
			\end{feynman}
			\end{tikzpicture}
			\end{gathered}
\hspace{2mm}
+2 \hspace{2mm}
		\begin{gathered}\begin{tikzpicture} \begin{feynman}[small, baseline=g1]
					\tikzfeynmanset{every vertex={dot,black,minimum size=1mm}}
				\vertex  (l1);
				\vertex [below =0.3cm of l1] (l2);
					\tikzfeynmanset{every vertex={empty dot,black,minimum size=1mm}}
				\vertex [right =0.3cm of l1] (r1) ;
				\vertex [below =0.3cm of r1] (r2);
				\diagram* {
					(l1) -- [] (r1),
					(l1) -- [] (r2),
				};	
			\end{feynman}\end{tikzpicture}\end{gathered}
\hspace{2mm} 
%
%
+2 \hspace{2mm}
		\begin{gathered}\begin{tikzpicture} \begin{feynman}[small, baseline=g1]
					\tikzfeynmanset{every vertex={dot,black,minimum size=1mm}}
				\vertex  (l1);
				\vertex [below =0.3cm of l1] (l2);
					\tikzfeynmanset{every vertex={empty dot,black,minimum size=1mm}}
				\vertex [right =0.3cm of l1] (r1) ;
				\vertex [below =0.3cm of r1] (r2);
				\diagram* {
					(r1) -- [] (l1),
					(r1) -- [] (l2),
				};	
			\end{feynman}\end{tikzpicture}\end{gathered}
\hspace{2mm} + \hspace{2mm}
			\begin{gathered}\begin{tikzpicture} \begin{feynman}[small, baseline=g1]
					\tikzfeynmanset{every vertex={dot,black,minimum size=1mm}}
				\vertex  (l1);
				\vertex [below =0.3cm of l1] (l2);
					\tikzfeynmanset{every vertex={empty dot,black,minimum size=1mm}}
				\vertex [right =0.3cm of l1] (r1) ;
				\vertex [below =0.3cm of r1] (r2);
				\diagram* {
					(l1) -- [] (l2),
					(r1) -- [] (r2),
				};	
			\end{feynman}\end{tikzpicture}\end{gathered}
\hspace{2mm} -4 \hspace{2mm}
			\begin{gathered}\begin{tikzpicture} \begin{feynman}[small, baseline=g1]
					\tikzfeynmanset{every vertex={dot,black,minimum size=1mm}}
				\vertex  (l1);
				\vertex [below =0.3cm of l1] (l2);
					\tikzfeynmanset{every vertex={empty dot,black,minimum size=1mm}}
				\vertex [right =0.3cm of l1] (r1) ;
				\vertex [below =0.3cm of r1] (r2);
				\diagram* {
					(l1) -- [] (r1),
					(l2) -- [] (r2),
				};	
			\end{feynman}\end{tikzpicture}\end{gathered} 
			\hspace{2mm}.
\end{align*}

The 2-loop zeros in entries (3,1,2), (3,2,1), (4,1,2) and (4,2,1) correspond to the mixing of a 6-point and a 4-point operator into a 4-point operator. 
In the off-shell calculation, there do exist non-zero two loop diagrams that may contribute to this mixing. These are of the form
\begin{equation}\label{eq:diagp2-example-twoinsertion}
    \ \text{$p$} \left\{  \hspace{-2mm}
    \begin{gathered}\begin{tikzpicture}\begin{feynman}[small]
    \vertex at (0.5,0);
    \tikzfeynmanset{every vertex={dot,black,minimum size=1.5mm}}
    \vertex (b3) at (0,0);
       \tikzfeynmanset{every vertex={dot,black,minimum size=1.5mm}}
    \vertex at (-0.14,-0.3) {\tiny $\O^{}_6$};
     \vertex at (1.25,-0.2) {\tiny $\O^{}_4$};
    \vertex [right =1cm of b3] (bb3);
    \vertex  at (0.3,0.17) {};
    \vertex at (-1,0.0) (aaaa3) {};
    \vertex at (-1,0.5) (aaaa4) {};
    \vertex at (-1,-0.5) (aaaa5) {};
    \vertex at (2,0) (eeee3){$p$};
    \diagram* {
    	(aaaa3) --[] (b3) -- [] (aaaa4),
    	(aaaa5) -- [] (b3),
    	(b3)-- [half left] (bb3) -- [half left] (b3),
    	(b3)-- [] (eeee3)
    };
    \end{feynman}\end{tikzpicture}\end{gathered}
    \right. \,.
\end{equation}
Such diagrams are proportional to $p^2$ (i.e.\ they vanish on shell), which after including all permutations require a counterterm proportional to $\partial^2\phi$. Such operators will be removed by field redefinitions. This diagram therefore does not contribute to the mixing into a 4-point operator in an on-shell basis.

All other zeros in the ADT can be explained by the absence of Feynman diagrams, which is captured by the second bound in  Eq.\ \eqref{boundconclusion}.
In conclusion, the corresponding zero in the ADT is not predicted by the second bound in Eq.\ \eqref{boundconclusion}, but requires the third bound. 

Theorem \ref{thmnew} is also consistent with the 1-loop results of Ref.\ \cite{Chala:2021pll} in the SM EFT.
However, there are many vanishing entries due to the absence of Feynman diagrams, which are not predicted by the theorem, because it is only based on the lengths of the involved operators. For the specific interactions of the SM EFT, the bounds can thus be improved. 
For instance, there are no non-scaleless 1-loop 1PI Feynman diagrams for the mixing of 
$\O_{\phi^6}= (\phi^\dagger\phi)^3$ and $\O^{(1)}_{\phi\psi_L}=(\phi^\dagger i \overset\leftrightarrow{D}_\mu\phi)(\overline{\psi}_L\gamma^\mu\psi_L)$
into $\O_{\phi^8}=(\phi^\dagger\phi)^4$. We leave improved  bounds for interactions with different types of particles to future work. 

\section{Discussion}
\label{s:disc}

Theorem~\ref{thmnew} gives a strong bound on the loop order below which certain entries of the ADT have to be zero. This includes orders at which non-zero off-shell Feynman diagrams with insertions of the considered operators do exist.
The theorem applies in minimal subtraction schemes of dimensional regularization for an arbitrary EFT without relevant couplings. In other theories with higher-spin particles, such as the SM EFT, further zeros exist that go beyond the limits of our theorem. 
That is because we analyze operators based only on the number of fields, being agnostic towards specific particle content and symmetries of a theory.
We further emphasize that Theorem~\ref{thmnew} assumes an on-shell basis; roughly this is any basis in which operators do not vanish in the on-shell limit as contact terms. In the case of a single operator insertion (i.e.\ linear order),  Theorem~\ref{thmnew}  reduces to the non-renormalization theorem given in Ref.~\cite{Bern:2019wie}. We thereby also sharpen the theorem of Ref.~\cite{Bern:2019wie}, clarifying that it only holds in on-shell bases.

We obtained explicit results for the anomalous dimension tensor at mass dimension eight in real and complex scalar EFT, using the off-shell methods laid out in \cite{Cao:2021cdt}. We verified that the ADT satisfies Theorem~\ref{thmnew}, pointing out the particular entries for which this is non-trivially true. 
We have thereby also delivered stringent tests of the theorem and the saturation of its limits. It thus appears there are no further zeros to be found in these theories, at least not in the basis of conformal primaries. This is to be contrasted with the case of the ADM in the linear mixing scenario studied in \cite{Cao:2021cdt}, where unexpected zeros were found in this basis. 
The explicit anomalous dimensions, in combination with the results of \cite{Cao:2021cdt},
provide new higher order data that can be used to analyze the flow at the Wilson-Fisher fixed point. We hope to return to this in a future study.

\section*{Acknowledgements}

We  thank J. Vermaseren for providing us with a private version of FORM which includes an interface to the graph generator by T. Kaneko. 
J.R.N.\ is grateful to the DESY Hamburg Theory group for hospitality. 

W.C. is also supported by the Global Science Graduate Course (GSGC) program of the University of Tokyo, and acknowledges support from JSPS KAKENHI Grants No. 19H05810 and No. 22J21553.  F. H. is supported by the NWO Vidi grant 680-47-551 and the UKRI FLF Mr/S03479x/1. 
T.M. is supported by the World Premier International Research Center Initiative (WPI) MEXT, Japan, and by JSPS KAKENHI grants JP19H05810, JP20H01896,  JP20H00153, and JP22K18712. J.R.N.\ is supported by the Deutsche Forschungsgemeinschaft (DFG, German Research Foundation) - Projektnummer 417533893/GRK2575 “Rethinking Quantum Field Theory”.

\bibliographystyle{JHEP}
\bibliography{refs}

\end{document}